\documentclass[aps,twocolumn,showpacs,amsmath,amssymb,prc]{revtex4-2}
\usepackage{graphicx,here}
\usepackage{multirow}
\usepackage{tikz}
\usepackage{epstopdf}
\usepackage[percent]{overpic}
\usepackage{scrextend}
\usepackage{xcolor,xspace}
\usepackage[ulem=normalem]{changes}
\usepackage{hyperref}
\hypersetup{colorlinks=True,urlcolor=blue,linkcolor=blue,citecolor=blue,filecolor=black}

\colorlet{Changes@Color}{red}
\usepackage{dcolumn,color,footnote,bm,braket}
\usepackage{url,longtable,tabularx}
\usepackage{threeparttable}

\usepackage{fancybox}
\usetikzlibrary{matrix}

\newcommand\+{\dagger}
\newcommand\jr{j_{\rho}}
\newcommand\mr{m_{\rho}}

\newcommand\mgtbeta{M({\mathrm{GT}})}
\newcommand\mfbeta{M({\mathrm{F}})}

\newcommand\hb{\hat{H}_{\mathrm{B}}}
\newcommand\hf{\hat{H}_{\mathrm{F}}}
\newcommand\hbf{\hat{V}_{\mathrm{BF}}}
\newcommand\hff{\hat{V}_{\nu\pi}}

\newcommand\db{\beta\beta}

\newcommand\znbb{0\nu\beta\beta}

\newcommand\ga{g_{\mathrm{A}}}
\newcommand\gv{g_{\mathrm{V}}}
\newcommand\ave{(g_{\mathrm{A}}/g_{\mathrm{V}})_\mathrm{eff}}

\newcommand\ft{\log{ft}}
\newcommand\btm{\beta^{-}}
\newcommand\btp{\beta^{+}}

\newcommand\vd{v_{\mathrm{d}}}

\newcommand\vt{v_{\mathrm{t}}}

\begin{document}

\title{$\beta$ decay and evolution of low-lying structure in Ge and As nuclei}

\author{Kosuke Nomura}
\email{knomura@phy.hr}
\affiliation{Department of Physics, Faculty of Science, 
University of Zagreb, HR-10000 Zagreb, Croatia}

\date{\today}

\begin{abstract}
A simultaneous calculation for the shape evolution 
and the related spectroscopic properties of the low-lying states, 
and the $\beta$-decay properties in the even- and odd-mass Ge and As nuclei 
in the mass $A\approx70-80$ region, within the framework 
of the nuclear density functional theory and the particle-core 
coupling scheme, is presented. 
The constrained self-consistent mean-field 
calculations using a universal energy density functional (EDF) 
and a pairing interaction determines 
the interacting-boson Hamiltonian for the even-even core nuclei, 
and the essential ingredients of the particle-boson 
interactions for the odd-nucleon 
systems, and of the Gamow-Teller and Fermi transition operators. 
A rapid structural evolution from $\gamma$-soft oblate to prolate 
shapes, as well as the spherical-oblate shape coexistence around the 
neutron sub-shell closure $N=40$, is suggested to occur in 
the even-even Ge nuclei. 
The predicted low-energy spectra, 
electromagnetic transition rates, and $\beta$-decay $\ft$ values 
are in a reasonable agreement with experiment. The predicted $\ft$ 
values reflect the structures of the wave functions for the 
initial and final nuclei of $\beta$ decay, which are, 
to a large extent, determined by the microscopic input 
provided by the underlying EDF calculation. 
\end{abstract}

\maketitle

\section{Introduction}

$\beta$ decay of the atomic nucleus is a weak-interaction 
process that converts protons into neutrons 
or vice versa, and is one of the important fundamental 
nuclear processes that not only helps to understand 
the structure of an individual nucleus, but 
is also essential for modeling astrophysical phenomena 
such as the nucleosynthesis of neutron-rich heavy elements. 
Experiments have been carried out at major radioactive-ion-beam 
facilities around the world, providing a wealth of new 
data on the $\beta$-decay half-lives of neutron-rich heavy nuclei 
\cite{dillmann2003,nishimura2011,quinn2012,lorusso2015,caballero2016}. 
From a theoretical point of view, calculation of the 
$\beta$ decay properties should be sensitive to the 
nature of the wave functions of the initial and final 
nuclei, and hence serves as a benchmark of theoretical models. 
A number of theoretical investigations for the $\beta$ 
decay have been made from various approaches such as 
the interacting boson and boson-fermion models (IBM and IBFM) 
\cite{navratil1988,DELLAGIACOMA1989,brant2004,yoshida2013,mardones2016,nomura2020beta-1,nomura2020beta-2,ferretti2020}, 
the quasi-particle random-phase approximations 
\cite{alvarez2004,sarriguren2015,boillos2015,pirinen2015,simkovic2013,mustonen2016,suhonen2017,ravlic2021}, 
and the nuclear shell model \cite{langanke2003,caurier2005,syoshida2018,suzuki2018}.

Precise measurements and theoretical investigations of 
the $\beta$-decay properties are also vital 
for determining matrix elements of 
double-$\beta$ ($\db$) decay, a process in which two 
successive $\beta$ decays occur between those nuclei with 
$(A,Z)$ and $(A,Z\pm2)$. 
Especially, the zero-neutrino mode of the $\db$ decay 
($\znbb$) is not allowed in the standard model of elementary 
particles, and the observation of this process would greatly 
advance current understandings of the electroweak 
fundamental symmetries \cite{avignone2008}.

The nuclei in the germanium (Ge) region around the 
neutron number $N=40$  
is among the challenging regions of the nuclear chart, 
and has been of great interests for recent theoretical 
\cite{niksic2014,kaneko2015,wang2015,nomura2017ge,budaca2019,awwad2020} 
and experimental studies 
\cite{toh2013,corsi2013,AYANGEAKAA2016,forney2018,minda2019,sekal2021}. 
Their low-lying states are characterized 
by a rich variety of nuclear structure phenomena, 
represented by a rapid shape evolution from one nucleus 
to another, which includes the emergence of the 
neutron $N=40$ subshell closure around $^{72}$Ge, 
the competition between multiple intrinsic shapes in 
the vicinity of the ground state within 
a single nucleus, i.e., shape coexistence 
\cite{heyde2011,AYANGEAKAA2016}, 
and the triaxial deformation around $^{76,78}$Ge 
\cite{toh2013,forney2018}. 
The nucleus $^{76}$Ge is of special importance, since 
it is a candidate nucleus as the $\znbb$ decay emitter, 
and its odd-odd neighbor 
$^{76}$As is considered a virtual intermediate 
state of the $\db$ decay.

In this paper, 
the $\beta$-decay rates and the evolution of low-lying 
collective structure of the even-$A$ and 
odd-$A$ Ge and the neighboring arsenic (As) isotopes 
in the $A\approx70-80$ region is investigated 
within a framework of the nuclear density functional theory 
and the particle-core coupling scheme. 
In this method, 
firstly the potential energy surfaces 
with the triaxial quadrupole shape 
degrees of freedom for the even-even nuclei are computed by means 
of the constrained self-consistent mean-field (SCMF) 
\cite{RS} calculations based on a universal energy density 
functional (EDF) and a pairing interaction. 
The low-lying structure of the even-even core nucleus 
is described by the IBM, 
with the Hamiltonian determined by mapping the SCMF 
energy surface onto the expectation value of the 
Hamiltonian \cite{nomura2008}. 
The particle-core coupling for the odd-$A$ and 
odd-odd systems is modelled within the frameworks of 
the IBFM \cite{iachello1979,IBFM} and the 
interacting boson-fermion-fermion 
models (IBFFM) \cite{brant1984,IBFM}, respectively. 
The essential ingredients 
of the particle-boson interaction terms, 
and of the Gamow-Teller and 
Fermi transition operators 
are determined by the same SCMF calculations. 
This theoretical procedure allows for a consistent 
calculation of nuclear 
$\beta$ decay and low-lying structure 
in a computationally feasible way, and has been used for studies 
of the $\beta$ decays of the odd-$A$ \cite{nomura2020beta-1} 
and even-$A$ \cite{nomura2020beta-2} Xe, Cs, Ba, 
and La regions with mass $A\approx130$, using as a 
microscopic input the Gogny-type EDF \cite{robledo2019}.

The paper is organized as follows. In Sec.~\ref{sec:theory} 
the theoretical framework, 
including the SCMF method, the procedure to construct 
the particle-core 
Hamiltonian, and the electromagnetic, Gamow-Teller, and 
Fermi transition operators, is described. 
In Sec.~\ref{sec:ee}, the calculated 
deformation energy surfaces and spectroscopic properties of 
the low-lying states of 
the even-even Ge nuclei, and the possible 
shape coexistence in $^{72}$Ge are discussed. 
The spectroscopic results for 
the odd-mass Ge and As, and the odd-odd As nuclei 
are shown in Secs.~\ref{sec:odd} and \ref{sec:doo}, 
respectively. 
In Sec.~\ref{sec:beta} 
the $ft$ values for the $\beta$ decays between the 
odd-mass and between the 
odd-odd nuclei are shown. Section~\ref{sec:summary} summarizes 
the main results.

\section{Theoretical framework\label{sec:theory}}

\subsection{Self-consistent mean-field calculations}

As the first step, the constrained SCMF calculations for 
the even-even $^{66-78}$Ge nuclei are performed within 
the relativistic Hartree-Bogoliubov (RHB) framework
\cite{vretenar2005,niksic2011,DIRHB} 
with the density-dependent point-coupling (DD-PC1) 
\cite{DDPC1} functional for the particle-hole channel, 
and a separable pairing force of 
finite range \cite{tian2009} 
for the particle-particle channel. 
The constraints imposed in the SCMF calculations 
are on the mass quadrupole moments, which are 
related to the polar deformation variables $\beta$ 
and $\gamma$ \cite{BM}. 
The constrained calculations produce 
the $(\beta,\gamma)$-deformation energy surfaces. 

\subsection{Particle-core Hamiltonian}

To compute the spectroscopic observables such as 
excitation spectra and transition rates, it is required 
to go beyond the static SCMF approximation, including 
the dynamical correlations arising from the restoration 
of broken symmetries and fluctuations in the 
collective coordinates \cite{RS}. 

The spectroscopic calculation is here carried out 
in terms of the IBM. In the following, 
the neutron-proton 
IBM (IBM-2) \cite{OAI} is used, because it is suitable to 
treat $\beta$ decay, in which 
both proton and neutron degrees of freedom 
should be explicitly considered. 
The IBM-2 consists of the neutron and proton 
monopole ($s_{\nu}$ and $s_{\pi}$), and 
quadrupole ($d_{\nu}$ and $d_{\pi}$) bosons. 
From a microscopic point of view \cite{OAIT,OAI}, 
the $s_{\nu}$ ($s_\pi$) and $d_\nu$ ($d_\pi$) bosons 
are associated with the 
collective $S_\nu$ ($S_\pi$) and $D_\nu$ ($D_\pi$) 
pairs of valence neutrons (protons) 
with angular momenta $J=0^{+}$ and $J=2^{+}$, 
respectively. Here the neutron (or proton) major 
oscillator shell
$N(\text{or}\;Z)=28-50$ is taken as the model space of the 
neutron (or proton) boson system. 
Hence for the $^{66-78}$Ge nuclei considered in this study, 
the number of the neutron bosons, 
$N_{\nu}$, varies within the range 
$2\leqslant N_\nu\leqslant 5$, while 
the number of the proton bosons is fixed, $N_\pi=2$. 

To deal with the even-even, odd-mass, and odd-odd nuclei 
simultaneously, both the collective (bosonic) and 
single-particle degrees of freedom are treated 
on the footing, within the neutron-proton IBFFM (IBFFM-2). 
The Hamiltonian of the IBFFM-2 is given by
\begin{align}
\label{eq:ham-ibffm2}
 \hat{H}=\hb + \hf^{\nu} + \hf^{\pi} + \hbf^{\nu} + \hbf^{\pi} + \hff,
\end{align}
where $\hb$ is the IBM-2 Hamiltonian
representing the 
bosonic even-even core, $\hf^{\nu}$ ($\hf^{\pi}$) 
is the single-neutron (proton) 
Hamiltonian, $\hbf^{\nu}$ ($\hbf^{\pi}$) represents 
the interaction between the odd neutron (proton) and 
the even-even IBM-2 core, 
and the last term $\hff$ is the residual 
neutron-proton interaction.

For the IBM-2 Hamiltonian the following form is employed:
\begin{align}
\label{eq:hb}
 \hb = 
&\epsilon_{d}(\hat{n}_{d_{\nu}}+\hat{n}_{d_{\pi}})
+\kappa\hat{Q}_{\nu}\cdot\hat{Q}_{\pi}
+ \kappa'\hat{L}\cdot\hat{L},
\end{align}
where in the first term, 
$\hat{n}_{d_\rho}=d^\+_\rho\cdot\tilde d_{\rho}$ 
($\rho=\nu$ or $\pi$) is the $d$-boson number operator, 
with $\epsilon_{d}$ the single $d$-boson
energy relative to the $s$-boson one, and 
$\tilde d_{\rho\mu}=(-1)^\mu d_{\rho-\mu}$. 
The second term stands for the quadrupole-quadrupole 
interaction between neutron and proton boson systems 
with strength $\kappa$, and 
$\hat Q_{\rho}=d_{\rho}^\+ s_{\rho} + s_{\rho}^\+\tilde d_{\rho} + \chi_{\rho}(d^\+_{\rho}\times\tilde{d}_{\rho})^{(2)}$ is the 
bosonic quadrupole operator, with the parameter $\chi_\rho$. 
The last term in Eq.~(\ref{eq:hb}) is a rotational 
term with strength $\kappa'$, where  
$\hat{L}=\sqrt{10}\sum_{\rho}(d^{\+}_{\rho}\times\tilde{d}_{\rho})^{(1)}$ 
is the bosonic angular momentum operator.

The single-nucleon Hamiltonian 
$\hf^{\rho}$ takes the form 
\begin{align}
\label{eq:hf}
 \hf^{\rho} = -\sum_{\jr}\epsilon_{\jr}\sqrt{2\jr+1}
  (a_{\jr}^\+\times\tilde a_{\jr})^{(0)}
\equiv
\sum_{\jr}\epsilon_{\jr}\hat{n}_{\jr},
\end{align}
where $\epsilon_{\jr}$ stands for the 
single-particle energy of the odd neutron $(\rho=\nu)$ 
or proton ($\rho=\pi$) orbital $\jr$. 
$a_{\jr}^{(\+)}$ represents 
particle annihilation (creation) operator, 
with $\tilde{a}_{\jr}$ defined by 
$\tilde{a}_{\jr\mr}=(-1)^{\jr -\mr}a_{\jr-\mr}$. 
On the right-hand side of Eq.~(\ref{eq:hf}), 
$\hat{n}_{\jr}$ stands for the number operator 
for the odd particle. 
For the fermion configuration space, 
the normal-parity orbitals 
$2p_{1/2}$, $2p_{3/2}$, and $1f_{5/2}$, 
for both neutron and proton are taken.

The boson-fermion interaction 
$\hbf^{\rho}$ here has a specific form \cite{IBFM}: 
\begin{equation}
\label{eq:hbf}
 \hbf^{\rho}
=\Gamma_{\rho}\hat{V}_{\mathrm{dyn}}^{\rho}
+\Lambda_{\rho}\hat{V}_{\mathrm{exc}}^{\rho}
+A_{\rho}\hat{V}_{\mathrm{mon}}^{\rho}. 
\end{equation}
The first, second, and third 
terms are dynamical quadrupole, 
exchange, and monopole interactions, respectively. 
By following the microscopic considerations with 
the generalized seniority scheme \cite{scholten1985,IBFM}, 
these are given as
\begin{align}
\label{eq:dyn}
&\hat{V}_{\mathrm{dyn}}^{\rho}
=\sum_{\jr\jr'}\gamma_{\jr\jr'}
(a^{\+}_{\jr}\times\tilde{a}_{\jr'})^{(2)}
\cdot\hat{Q}_{\rho'},\\
\label{eq:exc}
&\hat{V}^{\rho}_{\mathrm{exc}}
=-\left(
s_{\rho'}^\+\times\tilde{d}_{\rho'}
\right)^{(2)}
\cdot
\sum_{\jr\jr'\jr''}
\sqrt{\frac{10}{N_{\rho}(2\jr+1)}}
\beta_{\jr\jr'}\beta_{\jr''\jr} \nonumber \\
&{\quad}:\left(
(d_{\rho}^{\+}\times\tilde{a}_{\jr''})^{(\jr)}\times
(a_{\jr'}^{\+}\times\tilde{s}_{\rho})^{(\jr')}
\right)^{(2)}:
+ (\text{H.c.}),\\
\label{eq:mon}
&\hat{V}_{\mathrm{mon}}^{\rho}
=\hat{n}_{d_{\rho}}\hat{n}_{\jr},
\end{align} 
where the factors 
$\gamma_{\jr\jr'}=(u_{\jr}u_{\jr'}-v_{\jr}v_{\jr'})Q_{\jr\jr'}$, 
and $\beta_{\jr\jr'}=(u_{\jr}v_{\jr'}+v_{\jr}u_{\jr'})Q_{\jr\jr'}$ 
with 
$Q_{\jr\jr'}=\braket{\ell_{\rho}\frac{1}{2}\jr\|Y^{(2)}\|\ell'_\rho\frac{1}{2}\jr'}$ matrix element of the fermion 
quadrupole operator in the single-particle basis. 
$\hat{Q}_{\rho'}$ in Eq.~(\ref{eq:dyn}) is the same boson 
quadrupole operator as in the boson 
Hamiltonian (\ref{eq:hb}). 
The notation $:(\cdots):$ in Eq.~(\ref{eq:exc}) 
stands for normal ordering. 
Within the above formalism, the dynamical 
and exchange terms 
are dominated by the interactions between 
unlike particles, while the monopole term by 
like particle ones. 
In addition, the single-particle energy $\epsilon_{\jr}$ 
is replaced with the quasiparticle energy 
$\tilde\epsilon_{\jr}$.

For the residual 
neutron-proton interaction $\hff$ 
in Eq.~(\ref{eq:ham-ibffm2}), 
the following form \cite{brant1988} is adopted:
\begin{align}
\label{eq:hff}
\hff
=4\pi{\vd}
&\delta(\bm{r})
\delta(\bm{r}_{\nu}-r_0)
\delta(\bm{r}_{\pi}-r_0)
\nonumber\\
&+\vt
\left[
\frac{3({\bm\sigma}_{\nu}\cdot{\bf r})
({\bm\sigma}_{\pi}\cdot{\bf r})}{r^2}
-{\bm{\sigma}}_{\nu}
\cdot{\bm{\sigma}}_{\pi}
\right], 
\end{align}
where the first and second terms are $\delta$, 
and tensor interactions with strength 
parameters $\vd$, and $\vt$, respectively. 
Note that $\bm{r}=\bm{r}_{\nu}-\bm{r}_{\pi}$ 
and $r_0=1.2A^{1/3}$ fm.

\subsection{Procedure to build the Hamiltonian}

To construct the IBFFM-2 Hamiltonian 
(\ref{eq:ham-ibffm2}), first the IBM-2 Hamiltonian 
for the even-even core is determined. 
The parameters 
$\epsilon_d$, $\kappa$, $\chi_{\nu}$, and $\chi_{\pi}$ 
are determined by mapping the SCMF energy surface 
onto the expectation 
value of the Hamiltonian in the boson coherent state
\cite{dieperink1980,ginocchio1980}, 
so that the SCMF and IBM 
energy surfaces becomes similar to each other 
within the excitation energy of a few MeV 
with respect to the global minimum 
(see Refs.~\cite{nomura2008,nomura2010}, for details). 
The remaining parameter $\kappa'$ is fixed separately, 
in such a way \cite{nomura2011rot} 
that the cranking moment of inertia 
calculated in the intrinsic frame of the boson system 
at the global minimum becomes equal to the Inglis-Belyaev 
\cite{inglis1956,belyaev1961} value calculated by the RHB method. 
The derived IBM-2 parameters 
are listed in Table~\ref{tab:parab}. 

\begin{table}
\caption{\label{tab:parab}
Derived strength parameters for the 
IBM-2 Hamiltonian $\hb$ for the even-even $^{66-78}$Ge nuclei. 
}
 \begin{center}
 \begin{ruledtabular}
  \begin{tabular}{lccccc}
& $\epsilon_d$ & $\kappa$ & $\chi_{\nu}$ & $\chi_{\pi}$ & $\kappa'$ \\
Nucleus & (MeV) & (MeV) & & & (MeV)\\
\hline
$^{66}$Ge & 0.061 & $-0.680$ & $-0.07$ & $0.30$ & 0.015 \\ 
$^{68}$Ge & 0.091 & $-0.540$ & $0.20$ & $0.30$ & 0.051 \\ 
$^{70}$Ge & 0.535 & $-0.350$ & $0.80$ & $0.50$ & 0.069 \\ 
$^{72}$Ge & 0.747 & $-0.180$ & $0.30$ & $0.30$ & 0.050 \\ 
$^{74}$Ge & 0.950 & $-0.315$ & $0.30$ & $0.30$ & 0.000 \\ 
$^{76}$Ge & 0.600 & $-0.380$ & $-0.90$ & $-0.50$ & 0.000 \\ 
$^{78}$Ge & 0.620 & $-0.620$ & $-1.12$ & $-0.87$ & 0.000 \\ 
  \end{tabular}
 \end{ruledtabular}
 \end{center}
\end{table}

\begin{figure}[ht]
\begin{center}
\includegraphics[width=\linewidth]{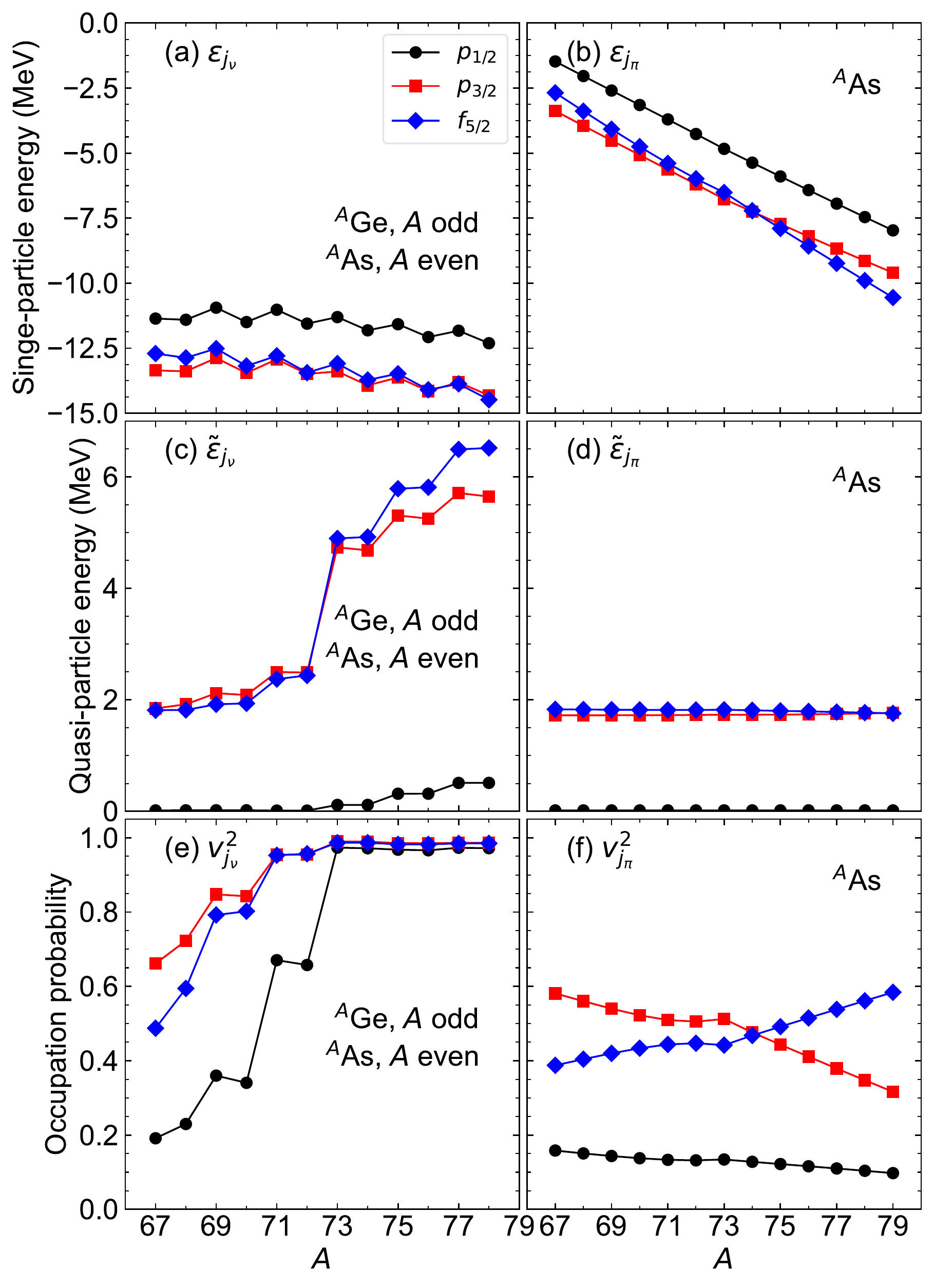}
\caption{The adopted spherical single-particle energies 
$\epsilon_{\jr}$ (a,b), 
quasiparticle energies $\tilde\epsilon_{\jr}$ (c,d), and 
occupation probabilities $v^2_{\jr}$ (e,f) 
for the $2p_{1/2}$, $2p_{3/2}$, 
and $1f_{5/2}$ orbitals for (left column) 
the odd neutron in the odd-$A$ Ge and even-$A$ As, 
and (right column) the odd proton in the odd-$A$ and even-$A$ As 
nuclei, calculated by the spherical RHB method.}
\label{fig:spe}
\end{center}
\end{figure}

Second, for each odd-mass nucleus 
the quasiparticle energies 
$\tilde\epsilon_{\jr}$ and occupation probabilities 
$v^{2}_{\jr}$ of the odd nucleons 
are calculated by the RHB method constrained to zero 
deformation $\beta=0$. 
These quantities are then 
used for the single-nucleon Hamiltonian $\hf^\rho$ 
(\ref{eq:hf}) and the boson-fermion interactions $\hbf^\rho$ 
(\ref{eq:hbf})--(\ref{eq:exc}).
Here, the fixed values of the strength parameters are adopted: 
$\Gamma_{\nu}=0.3$ MeV, $\Lambda_{\nu}=0.8$ MeV, 
and $A_{\nu}=-0.5$ MeV for the 
odd-$N$ Ge nuclei, and 
$\Gamma_{\pi}=0.3$ MeV, $\Lambda_{\pi}=0.3$ MeV, 
and $A_{\pi}=0$ MeV for the odd-$Z$ As. 
These values are determined so as to reasonably reproduce 
the experimental data on the low-energy negative-parity 
excitation spectra for the odd-mass nuclei.

Third, the same strength parameters 
$\{\Gamma_{\rho},\Lambda_{\rho},A_{\rho}\}$ as the ones 
employed for the neighboring odd-mass nuclei 
are used for the IBFFM-2 calculation on 
the odd-odd As ones. The quasiparticle energies 
and occupation probabilities are newly calculated. 
Finally, the fixed values of the 
strength parameters for the interaction 
$\hff$, $\vd=0.8$ MeV and $\vt=0.02$ MeV, are 
determined so that an overall reasonable agreement with 
the observed low-lying positive-parity states of 
the odd-odd As nuclei is obtained. 

Table~\ref{tab:comp} summarizes the even-even Ge core nuclei, 
and the neighboring odd-$N$ Ge, odd-$Z$ As, and 
odd-odd As nuclei considered in the present 
calculation. 
The $\epsilon_{\jr}$, $\tilde\epsilon_{\jr}$, 
and $v^2_{\jr}$ values of the odd nucleons, 
obtained from the spherical RHB calculations, 
are shown in Fig.~\ref{fig:spe}. 

\begin{table}
\caption{\label{tab:comp}
Even-even Ge core, neighboring odd-$N$ Ge, odd-$Z$ As, 
and odd-odd As nuclei considered in the present study. 
}
 \begin{center}
 \begin{ruledtabular}
  \begin{tabular}{cccc}
Core & odd-$N$  & odd-$Z$ & odd-odd  \\
\hline
$^{66}_{32}$Ge$_{34}$ & $^{67}_{32}$Ge$_{35}$ & $^{67}_{33}$As$_{34}$ & $^{68}_{33}$As$_{35}$ \\
$^{68}_{32}$Ge$_{36}$ & $^{69}_{32}$Ge$_{37}$ & $^{69}_{33}$As$_{36}$ & $^{70}_{33}$As$_{37}$ \\
$^{70}_{32}$Ge$_{38}$ & $^{71}_{32}$Ge$_{39}$ & $^{71}_{33}$As$_{38}$ & $^{72}_{33}$As$_{39}$ \\
$^{72}_{32}$Ge$_{40}$ & & $^{73}_{33}$As$_{40}$ & \\
$^{74}_{32}$Ge$_{42}$ & $^{73}_{32}$Ge$_{41}$ & $^{75}_{33}$As$_{42}$ & $^{74}_{33}$As$_{41}$ \\
$^{76}_{32}$Ge$_{44}$ & $^{75}_{32}$Ge$_{43}$ & $^{77}_{33}$As$_{44}$ & $^{76}_{33}$As$_{43}$ \\
$^{78}_{32}$Ge$_{46}$ & $^{77}_{32}$Ge$_{45}$ & $^{79}_{33}$As$_{46}$ & $^{78}_{33}$As$_{45}$ \\
  \end{tabular}
 \end{ruledtabular}
 \end{center}
\end{table}

\subsection{Electromagnetic transition operators}

The $E2$ operator $\hat T^{(E2)}$ 
in the IBFM-2 and IBFFM-2 takes the form \cite{IBFM}
\begin{align}
 \label{eq:e2}
\hat T^{(E2)}
= \hat T^{(E2)}_\text{B}
+ \hat T^{(E2)}_\text{F}
\end{align}
where the first and second terms are the 
boson and fermion parts, given respectively as
\begin{align}
 \label{eq:e2b}
\hat T^{(E2)}_\mathrm{B}
=\sum_{\rho=\nu,\pi}
e_\rho^\mathrm{B}\hat Q_\rho, 
\end{align}
and 
\begin{align}
 \label{eq:e2f}
\hat T^{(E2)}_\mathrm{F}
=-\frac{1}{\sqrt{5}}
&\sum_{\rho=\nu,\pi}
\sum_{\jr\jr'}
(u_{\jr}u_{\jr'}-v_{\jr}v_{\jr'})
\nonumber\\
&\times
\left\langle
\ell_\rho\frac{1}{2}\jr 
\bigg\| 
e^\mathrm{F}_\rho r^2 Y^{(2)} 
\bigg\|
\ell_\rho'\frac{1}{2}\jr'
\right\rangle
(a_{\jr}^\dagger\times\tilde a_{\jr'})^{(2)}.
\end{align}
The fixed values for the boson effective charges 
$e^\mathrm{B}_\nu = e^\mathrm{B}_\pi =0.0577$ $e$b 
are chosen so that the experimental 
$B(E2; 2^+_1\rightarrow 0^+_1)$ value 
for the well-deformed even-even core nucleus 
$^{72}$Ge is reproduced. 
The neutron and proton effective charges 
$e^\mathrm{F}_\nu =0.5$ $e$b 
$e^\mathrm{F}_\pi =1.5$ $e$b 
are adopted from the earlier IBFFM-2 
calculation on the odd-odd Cs nuclei \cite{nomura2020cs}. 
The $M1$ transition operator $\hat T^{(M1)}$ reads 
\begin{align}
 \label{eq:m1}
\hat T^{(M1)}
=\sqrt{\frac{3}{4\pi}}
&\sum_{\rho=\nu,\pi}
\Biggl[
g_\rho^\mathrm{B}\hat L_\rho
-\frac{1}{\sqrt{3}}
\sum_{\jr\jr'}
(u_{\jr}u_{\jr'}+v_{\jr}v_{\jr'})
\nonumber \\
&\times
\left\langle \jr \| g_l^\rho{\bf l}+g_s^\rho{\bf s} 
\| \jr' \right\rangle
(a_{\jr}^\+\times\tilde a_{\jr'})^{(1)}
\Biggr].
\end{align}
The empirical $g$ factors for the neutron and
proton bosons, $g_\nu^\mathrm{B}=0\,\mu_N$ and 
$g_\pi^\mathrm{B}=1.0\,\mu_N$, respectively, are adopted. 
For the neutron (or proton) $g$ factors, the standard 
Schmidt values $g_l^\nu=0\,\mu_N$ and $g_s^\nu=-3.82\,\mu_N$
($g_l^\pi=1.0\,\mu_N$ and $g_s^\pi=5.58\,\mu_N$) 
are used, with $g_s^\rho$ quenched by 30\% 
with respect to the free value.

\subsection{Gamow-Teller and Fermi transition operators}

The Gamow-Teller 
$\hat{T}^\mathrm{GT}$ and Fermi $\hat{T}^\mathrm{F}$ 
transition operators have the forms
\begin{align}
\label{eq:ogt}
&\hat{T}^{\rm GT}
=\sum_{j_{\nu}j_{\pi}}
\eta_{j_{\nu}j_{\pi}}^{\mathrm{GT}}
\left(\hat P_{j_{\nu}}\times\hat P_{j_{\pi}}\right)^{(1)}, \\
\label{eq:ofe}
&\hat{T}^{\rm F}
=\sum_{j_{\nu}j_{\pi}}
\eta_{j_{\nu}j_{\pi}}^{\mathrm{F}}
\left(\hat P_{j_{\nu}}\times\hat P_{j_{\pi}}\right)^{(0)}, 
\end{align}
respectively, with the coefficients 
\begin{align}
\label{eq:etagt}
\eta_{j_{\nu}j_{\pi}}^{\mathrm{GT}}
&= - \frac{1}{\sqrt{3}}
\left\langle
\ell_{\nu}\frac{1}{2}j_{\nu}
\bigg\|{\bm\sigma}\bigg\|
\ell_{\pi}\frac{1}{2}j_{\pi}
\right\rangle
\delta_{\ell_{\nu}\ell_{\pi}},\\
\label{eq:etafe}
\eta_{j_{\nu}j_{\pi}}^{\mathrm{F}}
&=-\sqrt{2j_{\nu}+1}
\delta_{j_{\nu}j_{\pi}}.
\end{align}
$\hat P_{\jr}$ in Eqs.~(\ref{eq:ogt}) and (\ref{eq:ofe}) 
is here identified as one of the one-particle creation 
operators 
\begin{subequations}
 \begin{align}
\label{eq:creation1}
&A^{\+}_{\jr\mr} = \zeta_{\jr} a_{{\jr}\mr}^{\+}
 + \sum_{\jr'} \zeta_{\jr\jr'} s^{\+}_\rho (\tilde{d}_{\rho}\times a_{\jr'}^{\+})^{(\jr)}_{\mr}
\\
\label{eq:creation2}
&B^{\+}_{\jr\mr}
=\theta_{\jr} s^{\+}_\rho\tilde{a}_{\jr\mr}
 + \sum_{\jr'} \theta_{\jr\jr'} (d^{\+}_{\rho}\times\tilde{a}_{\jr'})^{(\jr)}_{\mr},
\end{align}
and the annihilation operators
\begin{align}
\label{eq:annihilation1}
&\tilde{A}_{\jr\mr}=(-1)^{\jr-\mr}A_{\jr-\mr}\\ 
\label{eq:annihilation2}
&\tilde{B}_{\jr\mr}=(-1)^{\jr-\mr}B_{\jr-\mr}.  
\end{align}
\end{subequations}
Note that the operators in Eqs.~(\ref{eq:creation1}) 
and (\ref{eq:annihilation1}) 
conserve the boson number, whereas the ones 
in Eqs.~(\ref{eq:creation2}) and (\ref{eq:annihilation2}) 
do not. 
The $\hat{T}^{\rm GT}$ and $\hat{T}^{\rm F}$
operators are formed as a combination of two 
of the operators in 
(\ref{eq:creation1})--(\ref{eq:annihilation2}), 
depending on the type of the $\beta$ decay 
under study (i.e., $\beta^+$ or $\btm$) and 
on the particle or hole nature of bosons in 
the even-even IBM-2 core. 
It is also noted that the expressions in 
Eqs.~(\ref{eq:creation1})--(\ref{eq:annihilation2}) 
are of simplified forms of 
the most general one-particle transfer operators 
in the IBFM-2 \cite{IBFM}.

Within the generalized seniority scheme, 
the coefficients $\zeta_{j}$, $\zeta_{jj'}$, 
$\theta_{j}$, and $\theta_{jj'}$ 
in Eqs.~(\ref{eq:creation1}) and (\ref{eq:creation2}) 
can be given by \cite{dellagiacoma1988phdthesis}
\begin{subequations}
 \begin{align}
\label{eq:zeta1}
\zeta_{\jr}&= 
u_{\jr} \frac{1}{K_{\jr}'}, \\
\label{eq:zeta2}
\zeta_{\jr\jr'}
&= -v_{\jr} 
\beta_{\jr'\jr}
\sqrt{\frac{10}{N_{\rho}(2\jr+1)}}\frac{1}{K K_{\jr}'} , \\ 
\label{eq:theta1}
\theta_{\jr}
&= \frac{v_{\jr}}{\sqrt{N_{\rho}}} 
\frac{1}{K_{\jr}''},\\
\label{eq:theta2}
\theta_{\jr\jr'}
&= u_{\jr} 
\beta_{\jr'\jr}
\sqrt{\frac{10}{2\jr+1}} \frac{1}{K K_{\jr}''}. 
\end{align}
\end{subequations}
The factors $K$, $K_{\jr}'$, and $K_{\jr}''$ 
are defined as
\begin{subequations} 
\begin{align}
&K = \left( \sum_{\jr\jr'} 
\beta_{\jr\jr'}^{2} \right)^{1/2},\\
&K_{\jr}' = \left[ 1 + 2 
\left(\frac{v_{\jr}}{u_{\jr}}\right)^{2} \frac{\braket{(\hat 
n_{s_\rho}+1)\hat n_{d_\rho}}_{0^+_1}} {N_\rho(2\jr+1)} \frac{\sum_{\jr'} 
\beta_{\jr'\jr}^{2}}{K^{2}} \right]^{1/2} ,\\
&K_{\jr}'' = \left[ 
\frac{\braket{\hat n_{s_\rho}}_{0^+_1}}{N_\rho} 
+2\left(\frac{u_{\jr}}{v_{\jr}}\right)^{2} \frac{\braket{\hat 
n_{d_\rho}}_{0^+_1}}{2\jr+1} \frac{\sum_{\jr'} \beta_{\jr'\jr}^{2}}{K^{2}} 
\right]^{1/2},
\end{align} 
\end{subequations}
where $\hat n_{s_{\rho}}$ is the number operator 
for the $s_\rho$ boson and $\braket{\cdots}_{0^+_1}$ 
represents the expectation 
value of a given operator in the $0^+_1$ ground state 
of the even-even nucleus. 
In the expressions in Eqs.~(\ref{eq:zeta1}) to 
(\ref{eq:theta2}), the occupation 
$v_{\jr}$ and unoccupation $u_{\jr}$ amplitudes 
are the same as those used in the IBFM-2 (or IBFFM-2) 
calculations for the odd-mass (or odd-odd) nuclei. 
Within this framework, no additional phenomenological 
parameter is introduced for the GT and Fermi operators.

For a more detailed account on the formalism of the 
$\beta$-decay operators within the IBFM-2, 
the reader is referred to 
Refs.~\cite{dellagiacoma1988phdthesis,DELLAGIACOMA1989,IBFM}.

\begin{figure*}[ht]
\begin{center}
\includegraphics[width=.48\linewidth]{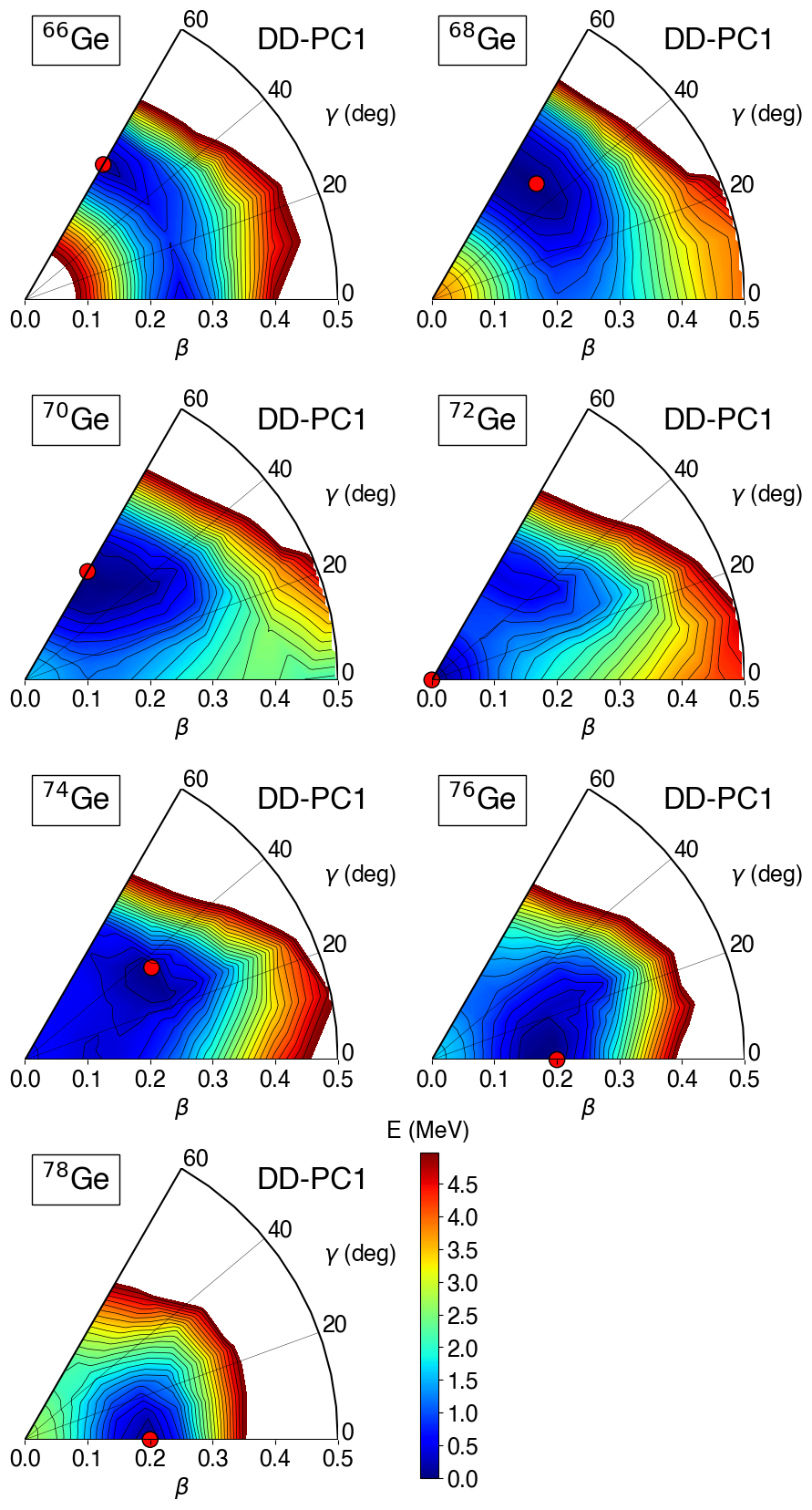}
\includegraphics[width=.48\linewidth]{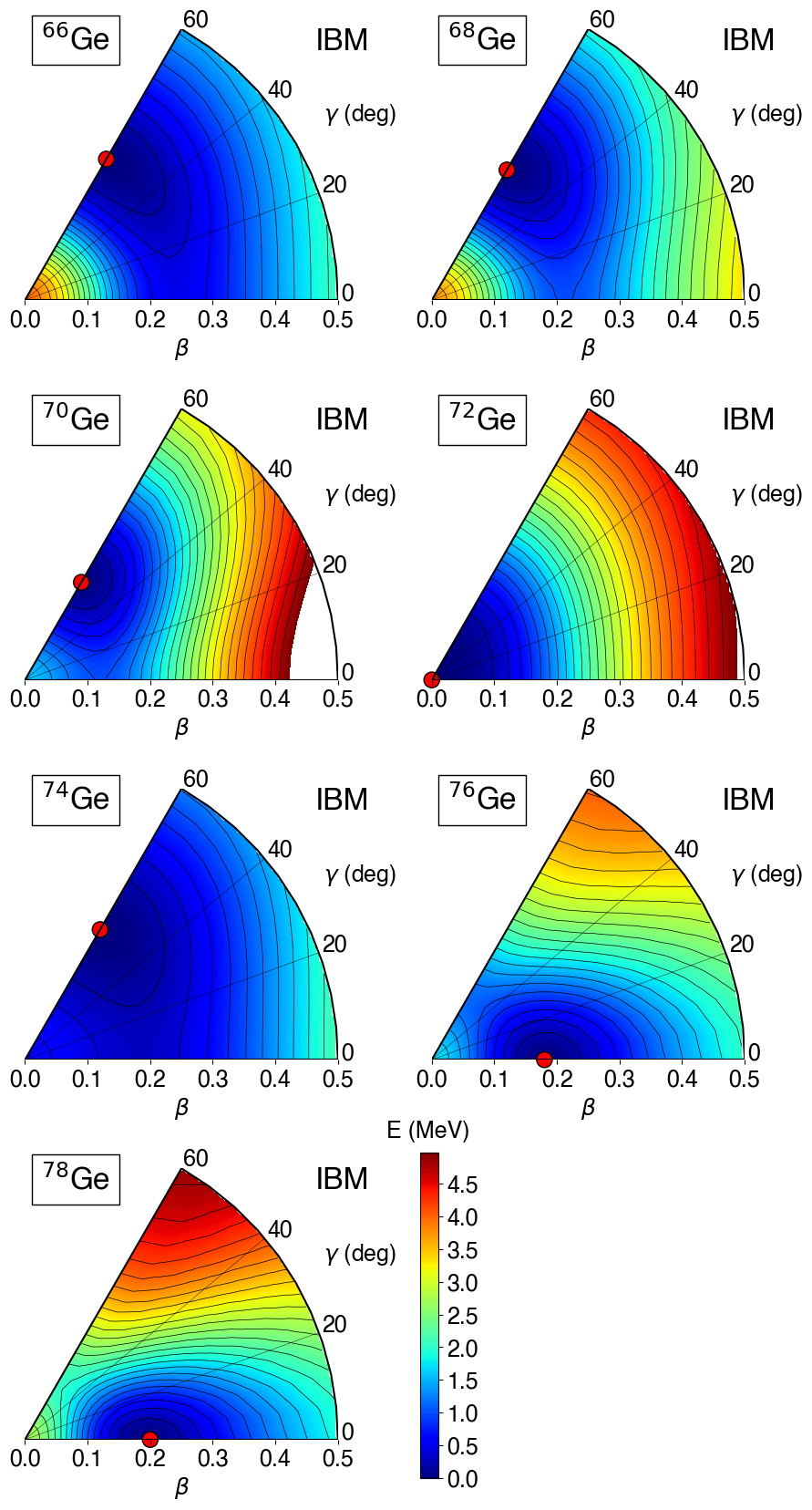}
\caption{SCMF and IBM $(\beta,\gamma)$-deformation 
energy surfaces for the even-even $^{66-78}$Ge nuclei. 
The energy difference between 
neighboring contours is 200 keV. The global minimum 
is identified by the solid circle. 
}
\label{fig:pes}
\end{center}
\end{figure*}

\section{Even-even nuclei\label{sec:ee}}

\subsection{Potential energy surfaces\label{sec:pes}}

In Fig.~\ref{fig:pes}, the contour plots of the SCMF 
quadrupole triaxial energy surfaces for the even-even 
$^{66-78}$Ge nuclei 
are shown as functions of 
the $(\beta,\gamma)$ deformations. 
The SCMF result indicates that the potential is 
generally soft in $\gamma$ deformation. 
The softness implies a substantial degree of 
shape mixing near the ground state. 
The SCMF energy surfaces shown in the figure suggests 
a transition from the $\gamma$-soft oblate ($^{66,68,70}$Ge), 
to spherical-oblate shape coexistence ($^{72}$Ge), 
to $\gamma$-soft ($^{74,76}$Ge), and to prolate shapes 
($^{78}$Ge). The appearance of the spherical 
ground-state minimum at $^{72}$Ge reflects the neutron 
$N=40$ sub-shell gap. 
A similar trend is obtained when another 
relativistic functional DD-ME2 \cite{DDME2} is used. 
The behavior of the energy surface as a function of 
the neutron number is also consistent with the one 
for the same even-even Ge nuclei, 
obtained in Ref.~\cite{nomura2017ge} with the 
Hartree-Fock-Bogoliubov calculations based on the 
Gogny-D1M \cite{D1M} EDF. 

The mapped IBM energy surfaces, 
shown also in Fig.~\ref{fig:pes}, basically have 
a similar topology in the vicinity of the global minimum 
and systematic trend with $N$ to those for the SCMF ones. 
A notable difference is that the IBM energy surface 
is rather flat in the region 
corresponding to large $\beta$ deformations, 
while the SCMF one becomes even more steeper. 
The difference arises mainly 
because the IBM-2 is built on the 
limited configuration space consisting of valence nucleon pairs, 
while the SCMF model is on all the constituent 
nucleons \cite{nomura2008,nomura2010}. 
It is also worth noting that the IBM-2 energy surface 
for $^{72}$Ge does not reproduce the local oblate minimum 
that is found in the SCMF counterpart. Within the IBM, 
such coexisting minima can be produced, e.g., by the inclusion 
of the configuration mixing between the normal and intruder 
states \cite{frank2004}. 
This point will be discussed in Sec.~\ref{sec:sc}. 

%
\begin{figure}
\begin{center}
\includegraphics[width=\linewidth]{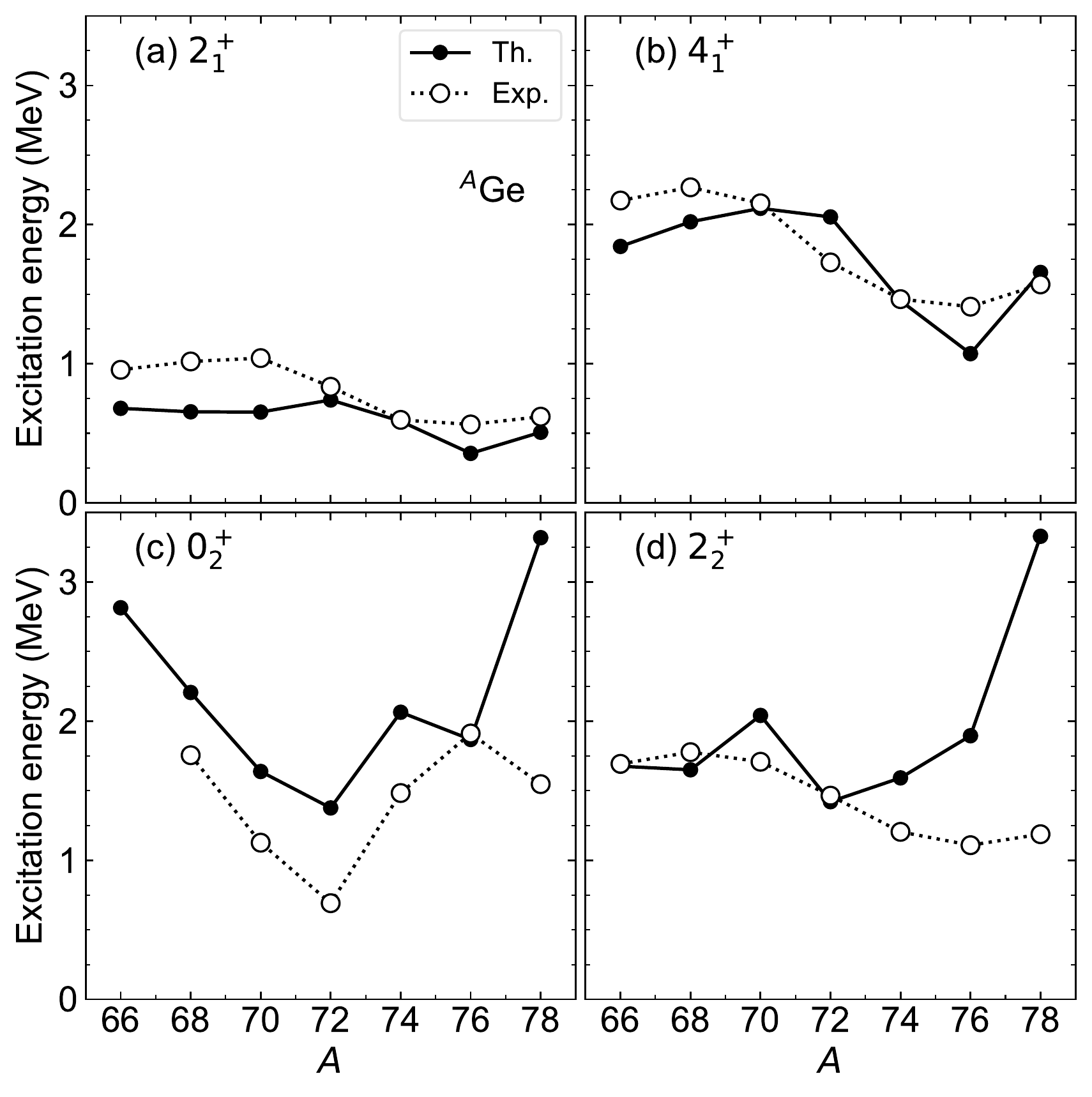}
\caption{Comparison of theoretical and experimental \cite{data}
excitation energies of the (a) $2^+_1$, (b) $4^+_1$, (c) $0^+_2$, 
and (d) $2^+_2$ states of the even-even $^{66-78}$Ge nuclei.}
\label{fig:evenge-level}
\end{center}
\end{figure}

%
\begin{figure}
\begin{center}
\includegraphics[width=\linewidth]{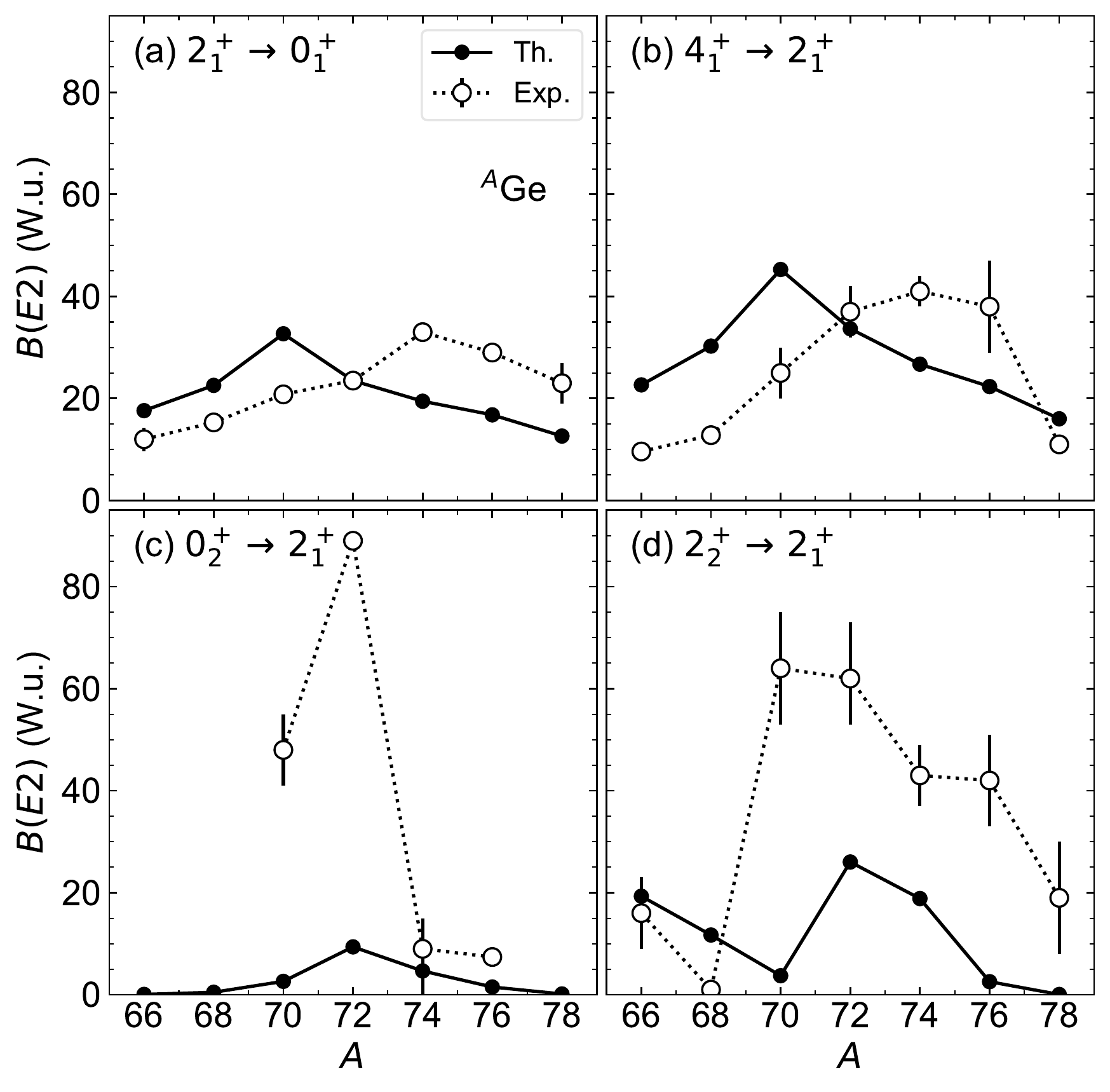}
\caption{Calculated and observed \cite{data} 
$B(E2)$ values for the 
transitions (a) $2^+_1\to0^+_1$, (b) $4^+_1\to2^+_1$, 
(c) $0^+_2\to2^+_1$, and (d) $2^+_2\to2^+_1$.}
\label{fig:evenge-e2}
\end{center}
\end{figure}

\subsection{Spectroscopic properties}

Figure~\ref{fig:evenge-level} shows evolution of the 
$2^+_1$, $4^+_1$, $0^+_2$, and $2^+_2$ energy levels for 
the considered even-even Ge, obtained from the diagonalization 
of the mapped IBM-2 Hamiltonian. 
In general, the calculation reproduces fairly well the 
observed systematics of the low-lying levels. 

For $^{66,68,70}$Ge, the calculation gives considerably lower 
$2^+_1$ energy level than the experimental one 
[see Fig.~\ref{fig:evenge-level}(a)]. 
This reflects the fact that the underlying SCMF energy surface 
for these nuclei shows a too pronounced deformation, and then 
the subsequent IBM-2 calculation produces a rather rotational-like 
energy spectrum characterized by the too compressed $2^+_1$ level. 
In addition, it should be kept in mind that in the present 
theoretical scheme the neutron-proton pair degrees of freedom 
are not taken into account both at the SCMF and IBM levels, 
which are supposed to play a certain 
role in those nuclei with $N\approx Z$ such as the ones 
considered here. 

In Fig.~\ref{fig:evenge-level}(c), 
the calculated $0^+_2$ excitation energies show a
similar overall behavior to the experimental counterparts. 
The calculation, however, systematically overestimates 
the data. 
Especially for $^{72}$Ge, the observed $0^+_2$ 
excitation energy is rather low ($\approx0.7$ MeV), 
while the mapped IBM-2 
overestimates it by a factor of 2. 
The emergence of the low-lying $0^+_2$ state is often 
attributed to the intruder excitation, which is not 
taken into account in the 
IBM-2 framework described above. 

The $2^+_2$ state is, in most of the cases, considered 
the bandhead of the $\gamma$-vibrational band. 
The mapped IBM-2 reproduces the experimental data 
up to $^{72}$Ge, but considerably overestimates the data 
for $^{76,78}$Ge. This is also due to the fact that the 
SCMF energy surface shows a pronounced deformation with 
a steep valley in both $\beta$ and $\gamma$ deformations. 
To reproduce this topology, the derived values 
for the parameters $\chi_\nu$ and $\chi_\pi$ 
for $^{76,78}$Ge have to have large negative values, 
as compared to the ones for $^{68-74}$Ge (see Table~\ref{tab:parab}). 
The resulting IBM-2 spectra for $^{76,78}$Ge 
are rather rotational like, where both the $\beta$- and 
$\gamma$-vibrational bands are generally high 
in energy with respect to the ground-state band. 
The other reason is that, as one approaches the neutron 
major shell closure $N=50$, the model space of the IBM, 
which is built on the finite number of valence nucleon pairs,  
becomes even smaller. For instance, 
there are only $N_{\nu}=N_{\pi}=2$ bosons for the nucleus $^{78}$Ge, 
which might not be large enough to describe satisfactorily 
the energy levels of the nonyrast states.

Figure~\ref{fig:evenge-e2} shows the calculated $B(E2)$ 
values for the transitions (a) $2^+_1\to0^+_1$, 
(b) $4^+_1\to2^+_1$, 
(c) $0^+_2\to2^+_1$, and (d) $2^+_2\to2^+_1$ for the 
even-even Ge nuclei, compared with the experimental 
data \cite{data}. 
The mapped IBM-2 gives maximal interband transition 
strengths in the ground-state 
band, $B(E2;2^+_1\to0^+_1)$ and $B(E2;4^+_1\to2^+_1)$, 
at $^{70}$Ge. This nucleus corresponds to the middle 
of the major shell $N=38$, at which the neutron 
boson number is maximal $N_\pi=5$, and thus the largest 
quadrupole collectivity is expected. 
The observed values for the above $B(E2)$ 
transition rates, however, show a peak around $^{74}$Ge. 
The calculated $B(E2;0^+_2\to2^+_1)$ and $B(E2;2^+_2\to2^+_1)$ 
values show a tendency similar to the observed one. 
The present IBM-2 calculation 
does not reproduce the large experimental 
$B(E2;0^+_2\to2^+_1)$ rates 
for $^{70,72}$Ge, mainly because it predicts 
for these nuclei a rather rotational-like spectrum, in which 
case the $0^+_2\to2^+_1$ transition is weak, 
and also because the model space does not include 
the effect of the configuration mixing between normal and 
intruder excitations.

\subsection{Shape coexistence\label{sec:sc}}

Shape coexistence 
is expected to occur in some of the considered 
even-even Ge nuclei. As an illustrative example, 
here the nucleus $^{72}$Ge is considered, for which the spherical 
global minimum and an oblate local minimum are 
suggested to occur in the corresponding 
SCMF quadrupole triaxial deformation energy map. 

A method to incorporate the effect of intruder configuration 
in the IBM-2 framework was proposed by Duval and Barrett 
\cite{duval1981}. In that method, two independent IBM-2 
Hamiltonians that correspond to the normal and intruder 
configurations are considered. In the case of proton 
two-particle-two-hole excitations, for instance, 
the intruder configuration is regarded as a system 
consisting of $N_\nu$ neutron and  $N_\pi+2$ proton bosons, 
under the assumption that the particle-like and hole-like 
bosons are not distinguished from each other. 
The two Hamiltonians are then admixed by a specific mixing 
interaction. 
The details about the configuration mixing IBM framework 
are found in Ref.~\cite{duval1981}, and its application 
to the Ge and Se region was made in Ref.~\cite{DUVAL1983}. 
The method of Duval and Barrett was also 
implemented in the mapped IBM framework, and 
the related applications to various mass regions, 
including the Ge and Se ones \cite{nomura2017ge}, 
were reported \cite{nomura2012sc,nomura2016sc,nomura2016zr,nomura2018cd}. 

Here it is assumed the proton $2p-2h$ excitations occur 
across the $Z=28$ major shell. The IBM-2 Hamiltonian 
configuration mixing (CM) is given by
\begin{align}
\label{eq:hcm}
 \hb^{\mathrm{CM}} = 
&{\hat{\cal{P}}}_{N_\pi}\hb^{N_\pi}{\hat{\cal{P}}}_{N_{\pi}}
\nonumber\\
&+{\hat{\cal{P}}}_{N_\pi+2}(\hb^{N_\pi+2}+\Delta){\hat{\cal{P}}}_{N_{\pi}+2}
+\hat V_{\mathrm{mix}},
\end{align}
where $\hb^{N_\pi+n}$ and ${\hat{\cal{P}}}_{N_\pi+n}$ $(n=0,2)$ 
are the Hamiltonian of and the projection operator onto 
the normal or intruder configuration space, respectively. 
$\Delta$ represents the energy needed to promote 
a proton boson across the $Z=28$ shell closure. 
The form of each unperturbed Hamiltonian $\hb^{N_\pi+n}$ 
is the same as the one in Eq.~(\ref{eq:hb}), but a specific 
three-body boson term
\begin{align}
 \kappa'' \sum_{\rho'\neq\rho}\sum_{L}
[d_\rho^\+\times d_\rho^\+\times d_{\rho'}^\+]^{(L)}\cdot
[\tilde d_{\rho'}\times \tilde d_\rho\times \tilde d_{\rho}]^{(L)}
\end{align}
is added to the Hamiltonian for the intruder configuration. 
The mixing interaction $\hat V_{\mathrm{mix}}$ in Eq.~(\ref{eq:hcm}) 
is given as
\begin{align}
 \hat V_{\mathrm{mix}} = 
\omega(s^\+_{\pi}\cdot s^\+_{\pi} + d^\+_{\pi}\cdot d^\+_{\pi}) + (\text{H.c.})
\end{align}
with $\omega$ mixing strength. 

The coherent state for the configuration-mixing IBM 
is given as the direct sum of the coherent state 
for each unperturbed configuration. The energy surface 
is, in general, expressed as the $2\times2$ 
coherent-state matrix \cite{frank2004}. 
Here the lower eigenvalue 
of the matrix is taken as the IBM energy surface. 
The parameters for each unperturbed Hamiltonian 
is determined by associating it to each mean-field 
minimum: the $0p-0h$ Hamiltonian for the spherical 
global 
minimum, 
and the $2p-2h$ one for the oblate local 
minimum for $^{72}$Ge. 
The off-set energy $\Delta$ and the mixing strength 
$\omega$ are determined so that the energy difference 
between the two mean-field minima and the barrier height 
for these minima are reproduced. 
The derived parameters are as follows. 
$\epsilon_d=1.6$ MeV, $\kappa=-0.22$ MeV, 
$\chi_\nu=0.30$, $\chi_\pi=0.30$, 
$\kappa'=\kappa''=0$ MeV
for the $0p-0h$ configuration, 
$\epsilon_d=0.9$ MeV, $\kappa=-0.21$ MeV, 
$\chi_\nu=0.30$, $\chi_\pi=0.22$, 
$\kappa'=0$ MeV, $\kappa''=0.11$ MeV 
for the $2p-2h$ configuration, $\Delta=1.82$ MeV, 
and $\omega=0.13$ MeV. 

%
\begin{figure}
\begin{center}
\includegraphics[width=\linewidth]{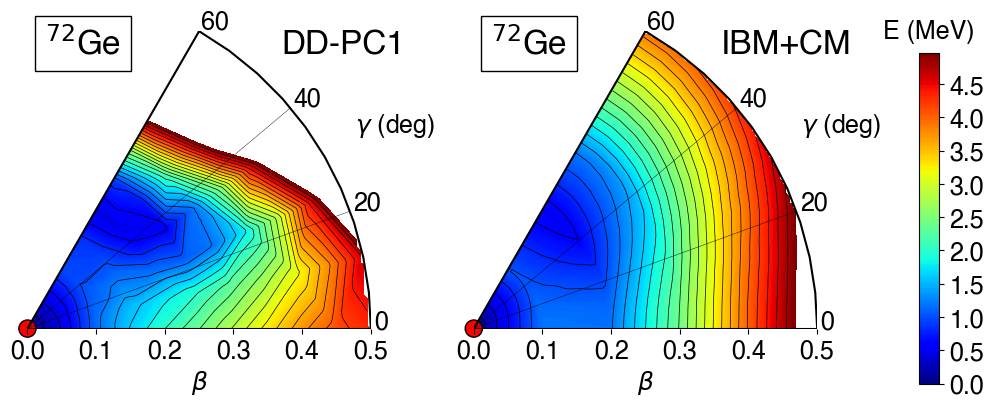}
\caption{Comparison of energy surfaces for $^{72}$Ge 
calculated with the SCMF method based on the DD-PC1 functional 
and mapped IBM-2 that includes configuration mixing (CM).}
\label{fig:pes-cm}
\end{center}
\end{figure}

%
\begin{figure}
\begin{center}
\includegraphics[width=.8\linewidth]{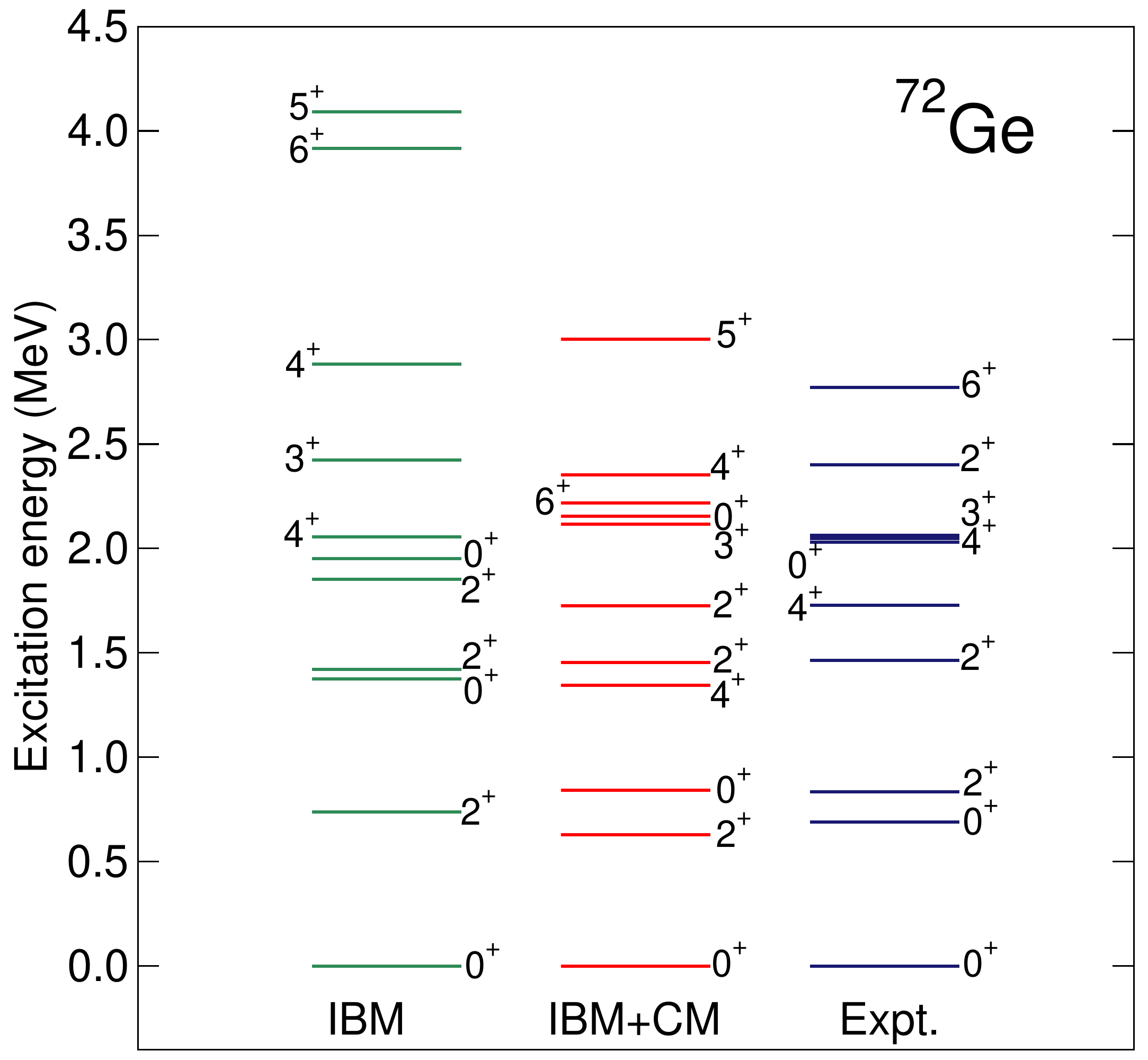}
\caption{Comparison of energy spectra of $^{72}$Ge 
resulting from the mapped IBM-2 with and without the 
configuration mixing (CM) with the experimental spectra.}
\label{fig:ibmcm}
\end{center}
\end{figure}

Figure~\ref{fig:pes-cm} compares the energy surface for the 
mapped IBM-2 that includes configuration mixing with the one obtained 
by the SCMF method. 
The IBM-2 surface now includes 
the oblate local minimum, consistent with the SCMF one. 
The resultant energy spectra in the cases where 
the configuration mixing is and 
is not performed are compared in Fig.~\ref{fig:ibmcm}. 
With the configuration mixing, the $0^+_2$ 
energy level is significantly lowered, being close 
to the observed $0^+_2$ level. 
In the configuration-mixing IBM-2 framework, 
the $E2$ transition operator is also extended as 
\begin{align}
 \hat T^{{(E2)}} = 
\sum_{n=0,2}\sum_{\rho}
{\hat{\cal{P}}}_{N_\pi+n}
e_{\mathrm{B},N_\pi+n}^{\rho}
\hat Q^{N_\pi+n}
{\hat{\cal{P}}}_{N_{\pi}+n}. 
\end{align}
If one uses the effective charges 0.05 $e$b and 0.06 $e$b 
for the normal and intruder configurations, respectively, 
the $B(E2; 0^+_2\to2^+_1)$ value for $^{72}$Ge is 
calculated to be 50 W.u. This is much greater than the 
value of 9.4 W.u., which is obtained without 
the configuration mixing, and is closer to the 
experimental data $89.0\pm1.5$ W.u. \cite{data}. 

The following discussion will be, however, based on the formalism 
without the configuration mixing and the higher-order 
boson terms, since the current versions of the IBFM and IBFFM 
codes do not handle these effects. 
It remains, therefore, an open question whether the configuration 
mixing, as well as the higher-order boson terms, plays a role 
in describing the low-lying states of the neighboring 
odd-mass and odd-odd nuclei, and the $\beta$-decay properties.

\section{Odd-$A$ Ge and As nuclei\label{sec:odd}}

%
\begin{figure}
\begin{center}
\includegraphics[width=\linewidth]{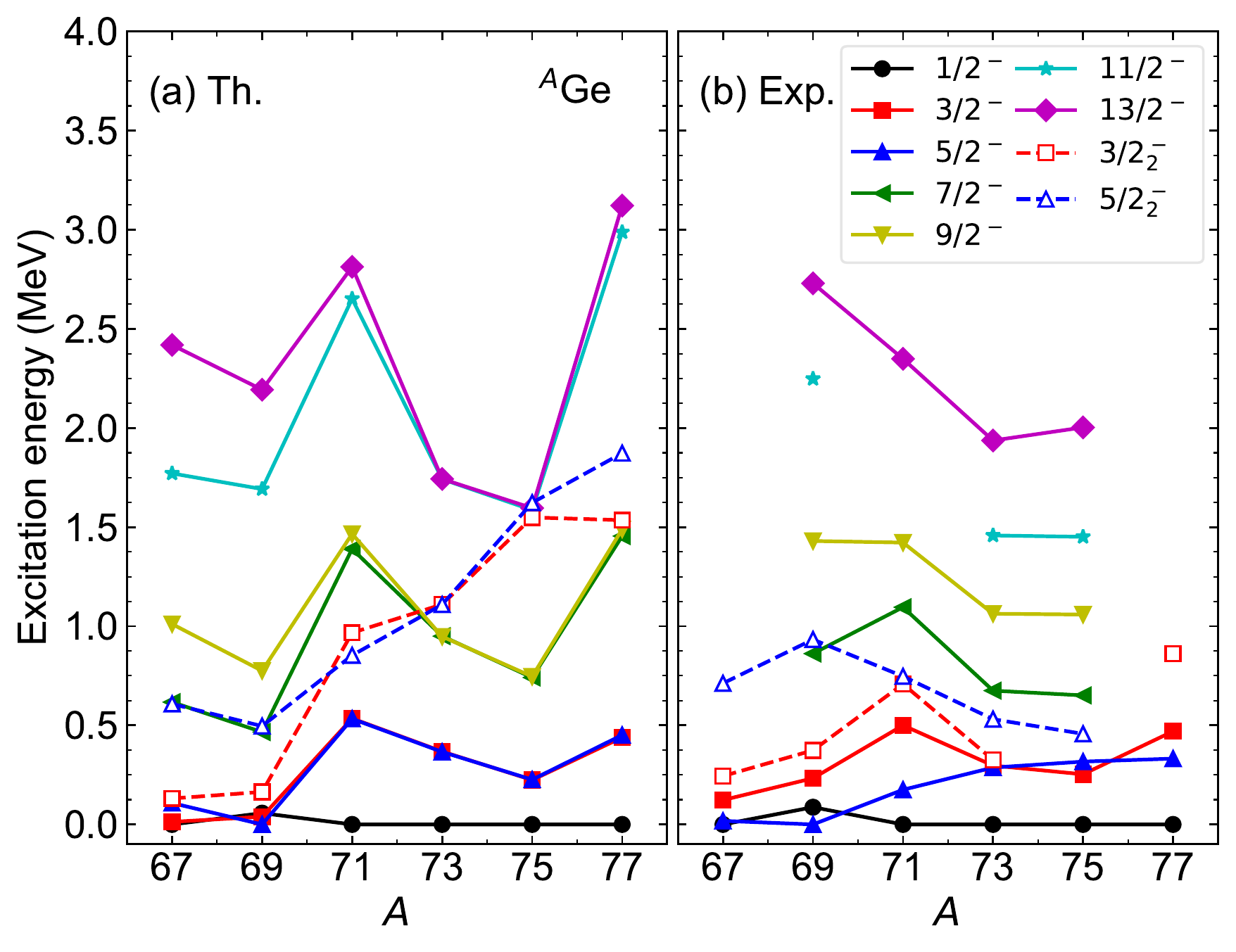}
\caption{Comparison of the calculated and experimental 
\cite{data}
low-lying negative-parity excitation spectra of the 
odd-$N$ Ge nuclei.}
\label{fig:oddge}
\end{center}
\end{figure}

Figure~\ref{fig:oddge} shows evolution of the low-energy  
negative-parity spectra of the odd-$N$ Ge isotopes as 
functions of the mass number $A$. Both the observed 
and predicted low-lying level structure near 
the ground state rapidly changes from $^{67}$Ge to $^{71}$Ge, 
represented by the change in the ground-state spin. 
There appears to be a significant structural change 
from $^{69}$Ge to $^{73}$Ge. This reflects the shape transition 
in the neighboring even-even core Ge nuclei, 
particularly the neutron 
$N=40$ sub-shell effect around $^{72}$Ge. 
As shown in Figs.~\ref{fig:spe}(c) and \ref{fig:spe}(e), 
the quasiparticle energies and occupation 
probabilities of the odd neutron also exhibit 
a sudden change from $^{71}$Ge to $^{73}$Ge. 

\begin{figure}
\begin{center}
\includegraphics[width=\linewidth]{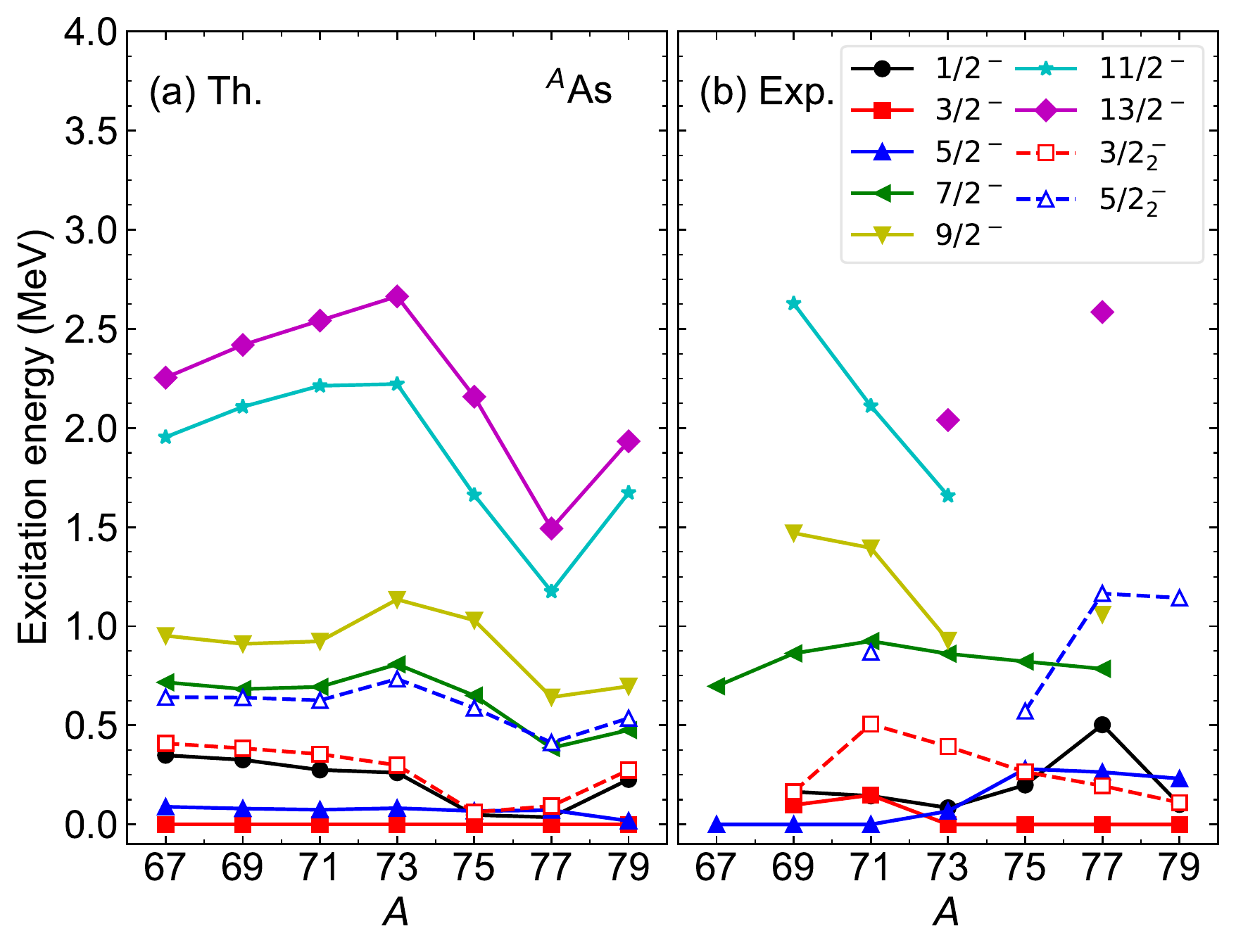}
\caption{Same as Fig.~\ref{fig:oddge}, but for 
the odd-$Z$ As nuclei.}
\label{fig:oddas}
\end{center}
\end{figure}

Figure~\ref{fig:oddas} gives the low-lying 
levels for the odd-$Z$ As nuclei. The predicted energy levels 
are more stable against the mass number $A$, especially in the 
region $A\leqslant73$, 
than in the case of the odd-$N$ Ge isotopes. 
One also notices that the ground-state spin is predicted 
to be $I={3/2}$ for all the As nuclei, which disagrees with 
the observed one $I=5/2^-$ for $^{67,69,71}$As. 
These features appear because the calculated $\tilde\epsilon_{j_\pi}$ 
and $v^2_{j_{\pi}}$ values for the proton orbitals 
only gradually change 
with $A$ [see Figs.~\ref{fig:spe}(d) and \ref{fig:spe}(f)], 
and also because of the use of the constant boson-fermion 
strengths. 
Similarly to the odd-$N$ Ge nuclei, a notable structural change 
in the low-lying levels is predicted to occur around 
$^{75,77}$As, which corroborates the rapid shape evolution 
in the even-even Ge core nuclei (see Fig.~\ref{fig:pes}). 
The situation is slightly different for the 
observed spectra, which suggest the change in the 
ground-state spin from $^{71}$As to $^{73}$As. 

%

%
\begin{figure}
\begin{center}
\includegraphics[width=\linewidth]{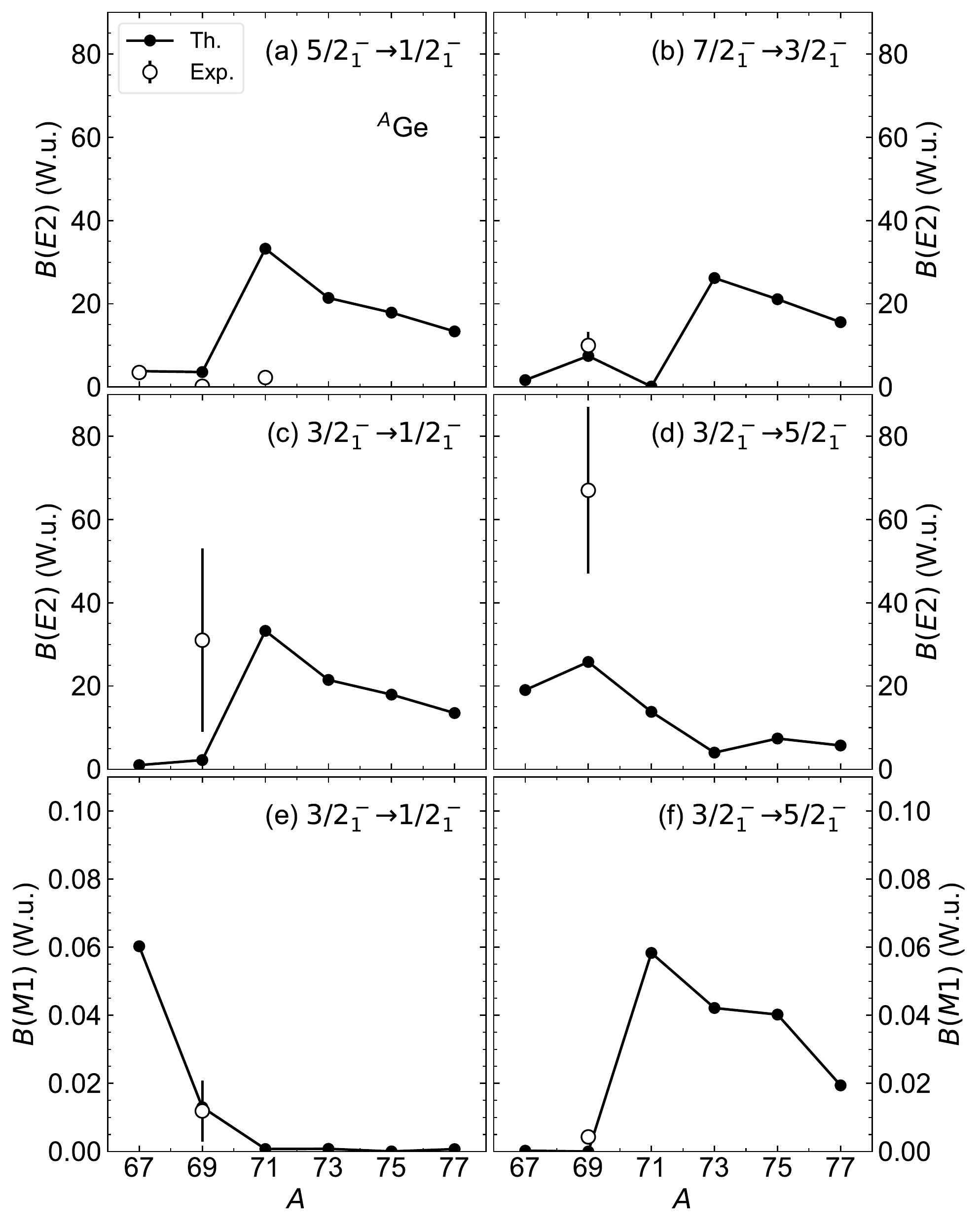}
\caption{Calculated $B(E2)$ and $B(M1)$ transition rates 
(in W.u.) between low-lying negative-parity states of odd-$N$ Ge 
isotopes. The available experimental data, taken from 
Ref.~\cite{data}, are also shown.}
\label{fig:oddge-em}
\end{center}
\end{figure}

%
\begin{figure}
\begin{center}
\includegraphics[width=\linewidth]{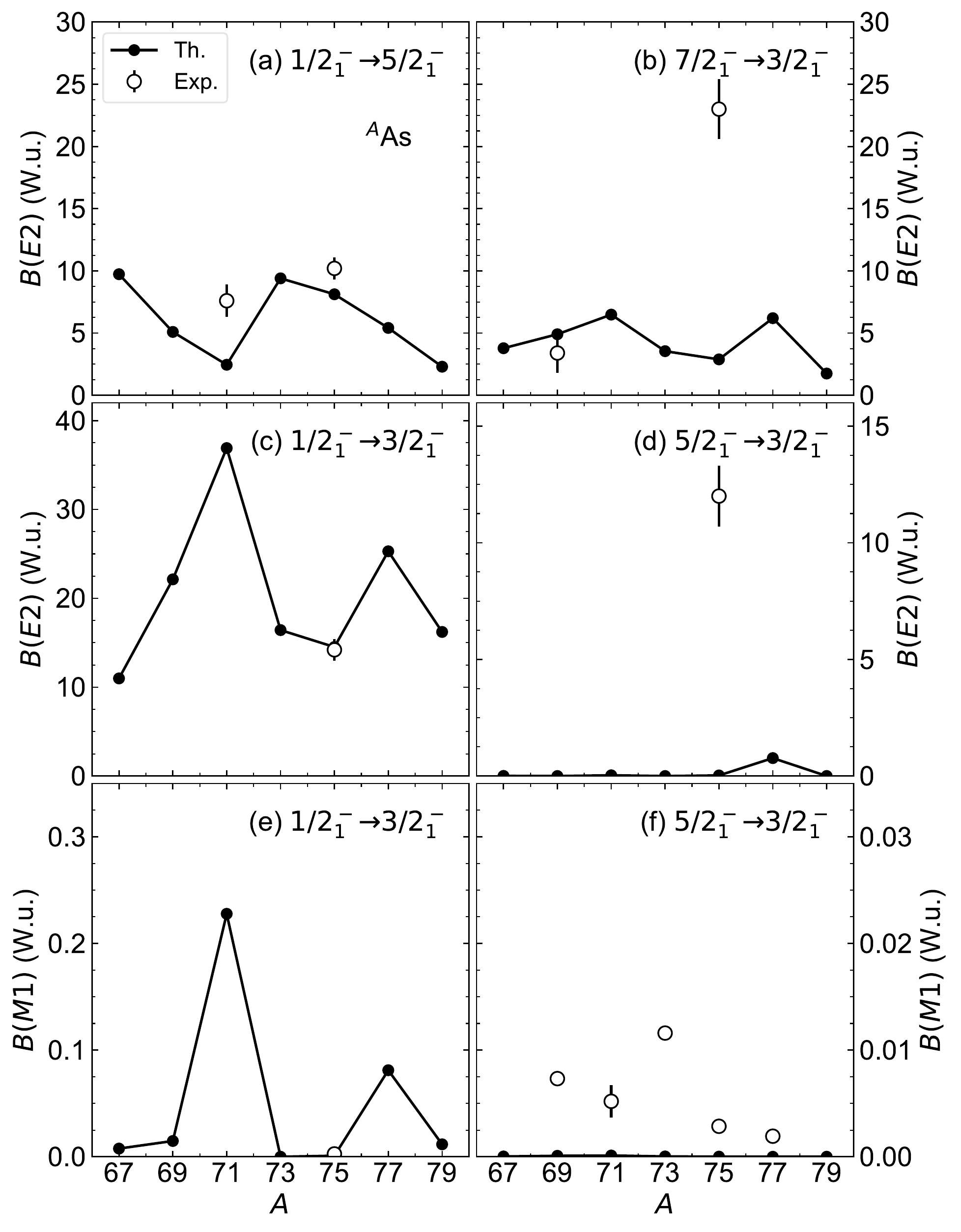}
\caption{Same as Fig.~\ref{fig:oddge-em}, but for the 
odd-$Z$ As nuclei.}
\label{fig:oddas-em}
\end{center}
\end{figure}

In Fig.~\ref{fig:oddge-em}, some predicted $B(E2)$ and 
$B(M1)$ transition strengths between the low-lying states 
for the odd-$N$ Ge are shown. Both the calculated $B(E2)$ 
and $B(M1)$ values show an abrupt change within 
the range $69\lesssim A\lesssim73$, indicating that the 
structure of the relevant IBFM-2 wave functions change. 
Note that the neighboring even-even 
Ge nuclei also undergoes rapid structural evolution 
between prolate and oblate shapes. 
In most cases, the calculated results shown 
in Fig.~\ref{fig:oddge-em} are in a reasonable agreement 
with the experimental data.

Figure~\ref{fig:oddas-em} shows 
the $B(E2)$ and $B(M1)$ transitions between low-lying 
negative-parity states of the odd-$Z$ As nuclei. 
One observes an irregular systematic in 
the predicted $B(E2;{1/2}^-_1\to{5/2}^-_1)$, 
$B(E2;{1/2}^-_1\to{3/2}^-_1)$, 
and $B(M1;{1/2}^-_1\to{3/2}^-_1)$ 
values. 
The last two quantities are particularly large at $^{71}$As. 
The occurrence of the irregularity indicates that 
the structure of the ${1/2}^-_1$ 
wave function obtained for $^{71}$As happens to be 
different from the ones for the neighboring nuclei. 
Furthermore, both the predicted $E2$ and $M1$ rates 
for the $5/2^-_1\to3/2^-_1$ transition 
are negligibly small, implying that the $3/2^-_1$ 
and $5/2^-_1$ have completely different structures 
in the IBFM-2. 

Table~\ref{tab:odd-mom} compares the calculated and experimental 
spectroscopic electric quadrupole $Q(I)$ and magnetic 
dipole $\mu(I)$ moments. In most cases, the present IBFM-2 
results are in a reasonable agreement with the observed 
$Q(I)$ and $\mu(I)$ moments, including the sign. 

\begin{table}
\caption{\label{tab:odd-mom}
Comparison of calculated and available experimental data for 
the electric quadrupole $Q(I)$ (in $e$b) and magnetic 
dipole $\mu(I)$ (in $\mu_N$) moments of the negative-parity 
states with spin $I$ for the odd-mass Ge and As 
nuclei. The data are taken from Ref.~\cite{stone2005}.
}
 \begin{center}
 \begin{ruledtabular}
  \begin{tabular}{lccc}
Nucleus 
& Moments & Th. & Exp. \\
\hline
$^{67}$Ge
& $\mu(5/2^-_{1})$ & $0.84$ & ${}$ \\
$^{69}$Ge
& $\mu(5/2^-_{1})$ & $0.79$ & $0.735\pm0.007$ \\
$^{71}$Ge
& $\mu(1/2^-_{1})$ & $0.45$ & $+0.547\pm0.005$ \\
& $\mu(5/2^-_{1})$ & $0.52$ & $+1.018\pm0.010$ \\
$^{73}$Ge
& $\mu(5/2^-_{1})$ & $0.76$ & ${}$ \\
$^{75}$Ge
& $\mu(1/2^-_{1})$ & $0.45$ & $+0.510\pm0.005$ \\
$^{77}$Ge
& $\mu(5/2^-_{1})$ & $1.22$ & ${}$ \\
$^{69}$As
& $\mu(5/2^-_{1})$ & $1.35$ & $+1.58\pm0.16$ \\
$^{71}$As
& $\mu(5/2^-_{1})$ & $1.31$ & $(+)1.674\pm0.02$ \\
& $Q(5/2^-_{1})$ & $0.004$ & $-0.017\pm0.010$ \\
$^{73}$As
& $\mu(5/2^-_{1})$ & $1.38$ & $+1.63\pm0.10$ \\
& $Q(5/2^-_{1})$ & $-0.02$ & $0.356\pm0.012$ \\
$^{75}$As
& $\mu(3/2^-_{1})$ & $2.85$ & $+1.43948\pm0.00007$ \\
& $Q(3/2^-_{1})$ & $-0.03$ & $+0.30\pm0.05$ \\
& $\mu(3/2^-_{2})$ & $0.91$ & $+1.0\pm0.2$ \\
& $\mu(5/2^-_{1})$ & $1.38$ & $+0.92\pm0.02$ \\
& $Q(5/2^-_{1})$ & $0.03$ & $0.30\pm0.10$ \\
$^{77}$As
& $\mu(3/2^-_{1})$ & $2.40$ & $+1.2946\pm0.0013$ \\
& $\mu(5/2^-_{1})$ & $1.35$ & $+0.74\pm0.02$ \\
& $Q(5/2^-_{1})$ & $0.06$ & $<0.75$ \\
  \end{tabular}
 \end{ruledtabular}
 \end{center}
\end{table}

%
\begin{figure}
\begin{center}
\includegraphics[width=\linewidth]{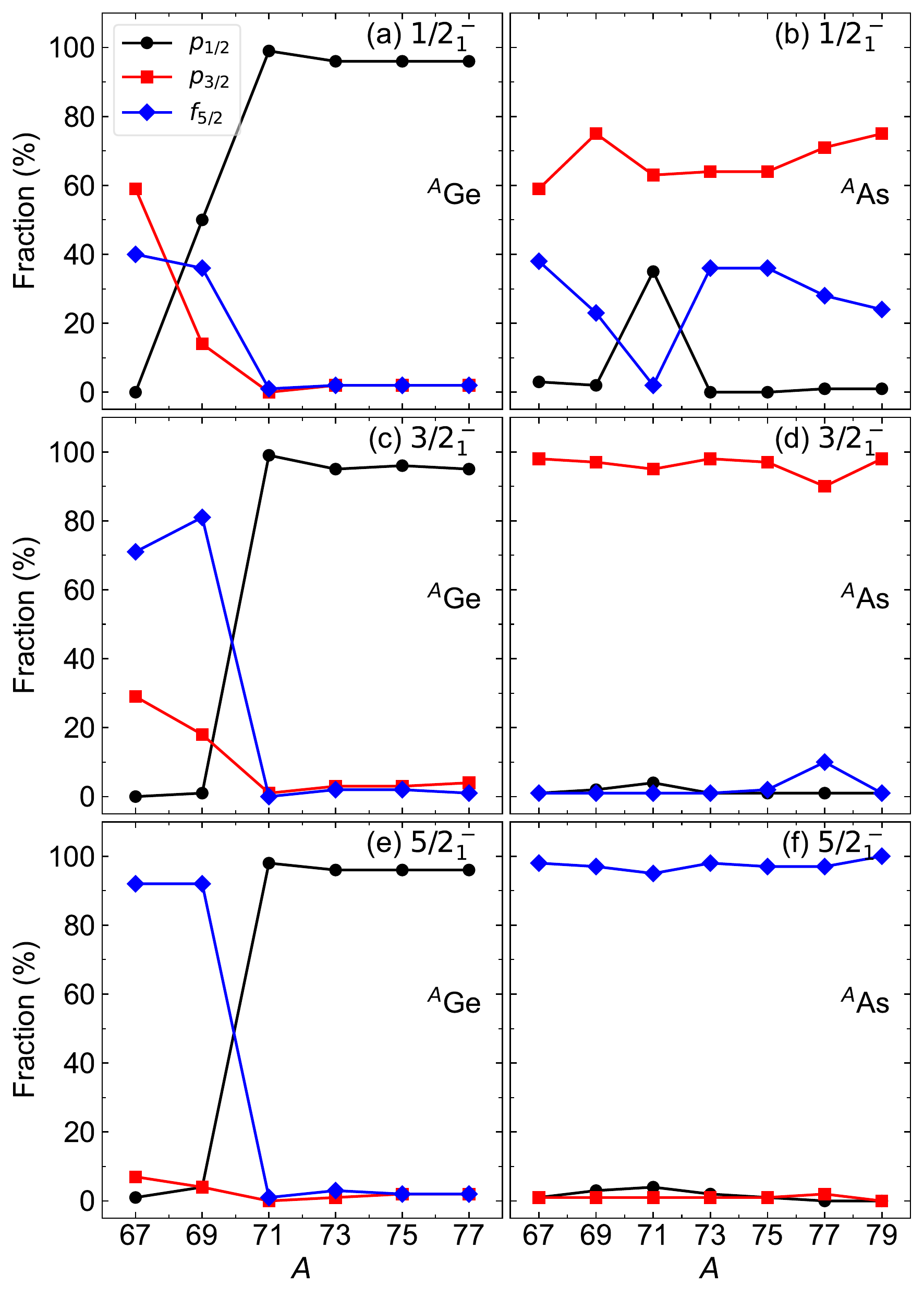}
\caption{Fractions of the $2p_{1/2}$, 
$2p_{3/2}$, and $1f_{5/2}$ single-particle configurations 
in the IBFM-2 wave functions for the ${1/2}^-_1$, 
${3/2}^-_1$, and ${5/2}^-_1$ states of the odd-$N$ Ge 
and odd-$Z$ As nuclei. }
\label{fig:odd-wf}
\end{center}
\end{figure}

To help interpreting the nature of the low-lying states 
in the considered odd-$A$ systems, Fig.~\ref{fig:odd-wf} shows 
the fractions of the $2p_{1/2}$, $2p_{3/2}$, 
and $1f_{5/2}$ single-particle 
configurations in the IBFM-2 wave functions for the ${1/2}^-_1$, 
${3/2}^-_1$, and ${5/2}^-_1$ states of the odd-$N$ Ge 
and odd-$Z$ As nuclei. 
For $^{67,69}$Ge, two or three single-particle 
configurations make sizable ($>10$ \%) contributions 
to the ${1/2}^-_1$ and ${3/2}^-_1$ wave functions. 
For the Ge nuclei with $A\geqslant71$, 
the three states ${1/2}^-_1$, ${3/2}^-_1$, and ${5/2}^-_1$ 
are almost entirely made of the neutron $2p_{1/2}$ 
configurations. This is evident from Fig.~\ref{fig:spe}(c), 
in which there is a larger energy gap 
between the $2p_{1/2}$ and the $2p_{3/2}$ and $1f_{5/2}$ 
quasi-neutron energies for the odd-$A$ Ge with 
$A\geqslant73$ than for the ones with $A\leqslant71$. 
The ${3/2}^-_1$ and ${5/2}^-_1$ states 
of the odd-$Z$ As nuclei are mostly accounted for by 
the proton $2p_{3/2}$ and $1f_{5/2}$ configurations, 
respectively. This explains the vanishing 
$B(E2;5/2^-_1\to3/2^-_1)$ and $B(M1;5/2^-_1\to3/2^-_1)$ 
rates, shown in Figs.~\ref{fig:oddas-em}(d) and 
\ref{fig:oddas-em}(f). 
As seen in Fig.~\ref{fig:odd-wf}(b), 
for the odd-$Z$ As nuclei 
the largest contribution 
($\approx60$ \%) to the ${1/2}^-_1$ wave function is 
from the $2p_{3/2}$ configuration, 
while either the $2p_{1/2}$ or $1f_{5/2}$ configuration 
constitutes about 30 \% of the wave function. 
Also, the ${1/2}^-_1$ wave function for $^{71}$As 
has a different composition from the ones for 
the neighboring isotopes, 
and this explains the irregular behavior of those 
calculated $B(E2)$ and $B(M1)$ values that involve 
the ${1/2}^-_1$ state (see Figs.~\ref{fig:oddas-em}(a), 
\ref{fig:oddas-em}(c), and \ref{fig:oddas-em}(e)). 

%
\begin{figure}
\begin{center}
\includegraphics[width=\linewidth]{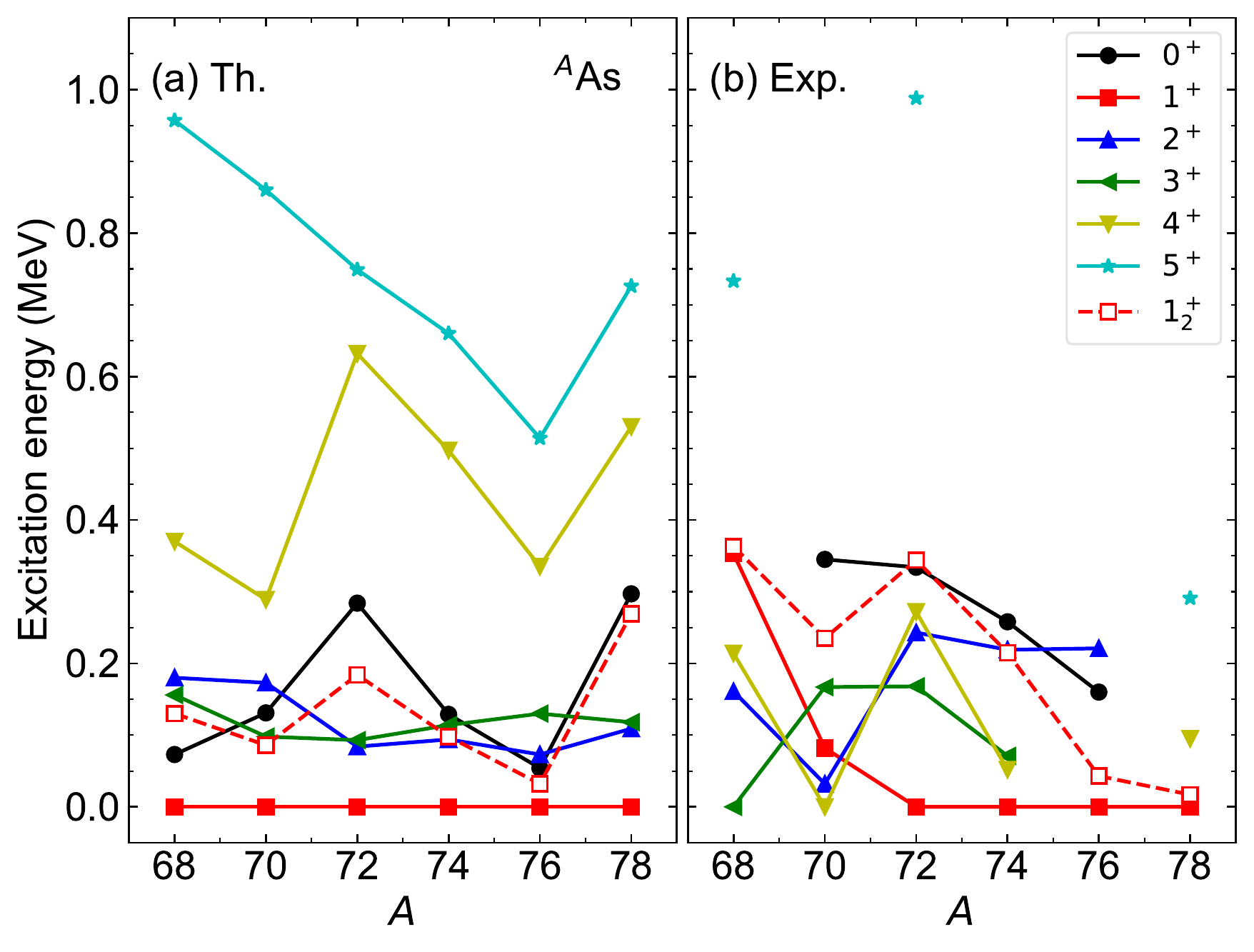}
\caption{Same as Fig.~\ref{fig:oddge}, but for the 
low-energy positive-parity states of odd-odd As nuclei.}
\label{fig:dooas}
\end{center}
\end{figure}

\section{Odd-odd As nuclei\label{sec:doo}}

Figure~\ref{fig:dooas} shows the results for the low-energy 
positive-parity spectra for the odd-odd As nuclei, obtained 
with the IBFFM-2. For those nuclei with mass $A\geqslant72$, 
the observed ground-state spin of $I=1^+$ is reproduced. 
The corresponding IBFFM-2 $1^+_1$ wave functions for these 
nuclei are dominated (approximately 80 \%) by the neutron-proton 
pair component 
$[\nu{p_{1/2}}\otimes\pi{p_{3/2}}]^{J=1^+}$ 
coupled to the even-even boson core nuclei.
This configuration plays a less important role 
in the predicted $1^+_1$ ground states for $^{68,70}$As.  
Instead, several configurations 
are admixed in these nuclei, with the major contributions 
coming from the components 
$[\nu{f_{5/2}}\otimes\pi{p_{3/2}}]^{J=3^+}$ (40 \%), 
$[\nu{p_{3/2}}\otimes\pi{p_{3/2}}]^{J=2^+}$ (16 \%), 
and $[\nu{f_{5/2}}\otimes\pi{p_{3/2}}]^{J=1^+}$ (13 \%) 
for $^{68}$As, and 
$[\nu{f_{5/2}}\otimes\pi{p_{3/2}}]^{J=1^+}$ (44 \%), 
$[\nu{p_{1/2}}\otimes\pi{p_{3/2}}]^{J=1^+}$ (12 \%), and 
$[\nu{f_{5/2}}\otimes\pi{p_{3/2}}]^{J=3^+}$ (10 \%) 
for $^{70}$As. 

Here the IBFFM-2 does not reproduce for $^{68}$As and $^{70}$As 
the observed ground state spins of $I=3^+$ and $4^+$, respectively. 
For $^{68}$As, main contributions to the IBFFM-2 wave function 
of the $3^+_1$ state come from the pair components 
$[\nu{f_{5/2}}\otimes\pi{p_{3/2}}]^{J=3^+}$ (70 \%), and 
$[\nu{f_{5/2}}\otimes\pi{f_{5/2}}]^{J=3^+}$ (11 \%). 
For both $^{68}$As and $^{70}$As, 
$[\nu{f_{5/2}}\otimes\pi{f_{5/2}}]^{J=4^+}$ 
constitutes approximately 75 \% of the $4^+_1$ wave function. 
For the $A\geqslant72$ Ge nuclei, 
the configuration $[\nu{p_{1/2}}\otimes\pi{f_{5/2}}]^{J=3^+}$  
accounts for $\approx70$ \% of the wave functions for 
both the $3^+_1$ and $4^+_1$ states. 

Table~\ref{tab:doo-mom} lists the calculated $Q(I)$ 
and $\mu(I)$ moments for the odd-odd As nuclei, 
as compared with the available experimental data 
\cite{stone2005}. 
The observed moments, especially their sign, are 
reasonably reproduced by the IBFFM-2. 
\begin{table}
\caption{\label{tab:doo-mom}
Same as Table~\ref{tab:odd-mom}, but for the positive-parity 
states of odd-odd As nuclei. 
}
 \begin{center}
 \begin{ruledtabular}
  \begin{tabular}{lccc}
Nucleus 
& Moments & Th. & Exp. \\
\hline
$^{68}$As
& $Q(3^+_{1})$ & $0.02$ & ${}$ \\
& $\mu(3^+_{1})$ & 2.13 & ${}$ \\
$^{70}$As
& $Q(4^+_{1})$ & $0.10$ & $+0.09\pm0.02$ \\
& $\mu(4^+_{1})$ & 1.72 & $+2.1061\pm0.0002$ \\
$^{72}$As
& $Q(3^+_{1})$ & $0.07$ & ${}$ \\
& $\mu(1^+_{1})$ & 1.92 & ${}$ \\
& $\mu(3^+_{1})$ & 1.38 & $+1.58\pm0.02$ \\
$^{74}$As
& $\mu(1^+_{1})$ & 2.10 & ${}$ \\
& $\mu(4^+_{1})$ & 1.94 & $+3.24\pm0.04$ \\
$^{76}$As
& $\mu(1^+_{1})$ & 1.71 & $+0.559\pm0.005$ \\
$^{78}$As
& $\mu(1^+_{1})$ & 2.44 & ${}$ \\
  \end{tabular}
 \end{ruledtabular}
 \end{center}
\end{table}

\section{$\beta$ decay\label{sec:beta}}

The $ft$ values for the $\beta$ 
decays of the odd-$A$ As into Ge nuclei, 
and of the even-$A$ As into Ge nuclei, and vice versa, 
are computed by the formula
\begin{align}
\label{eq:ft}
ft=\frac{K}{|\mfbeta|^2+\left(\frac{\ga}{\gv}\right)^2|\mgtbeta|^2},
\end{align}
where the constant $K=6163$ sec, and 
$\mfbeta$ and $\mgtbeta$ are the reduced matrix elements 
of the Fermi $\hat{T}^{\mathrm{F}}$ (\ref{eq:ofe}) 
and Gamow-Teller $\hat{T}^{\mathrm{GT}}$ (\ref{eq:ogt}) 
operators, respectively. 
The free values of the vector and axial vector 
coupling constants $\gv=1$ and $\ga=1.27$, respectively, 
are used.

\begin{figure}[ht]
\begin{center}
\includegraphics[width=\linewidth]{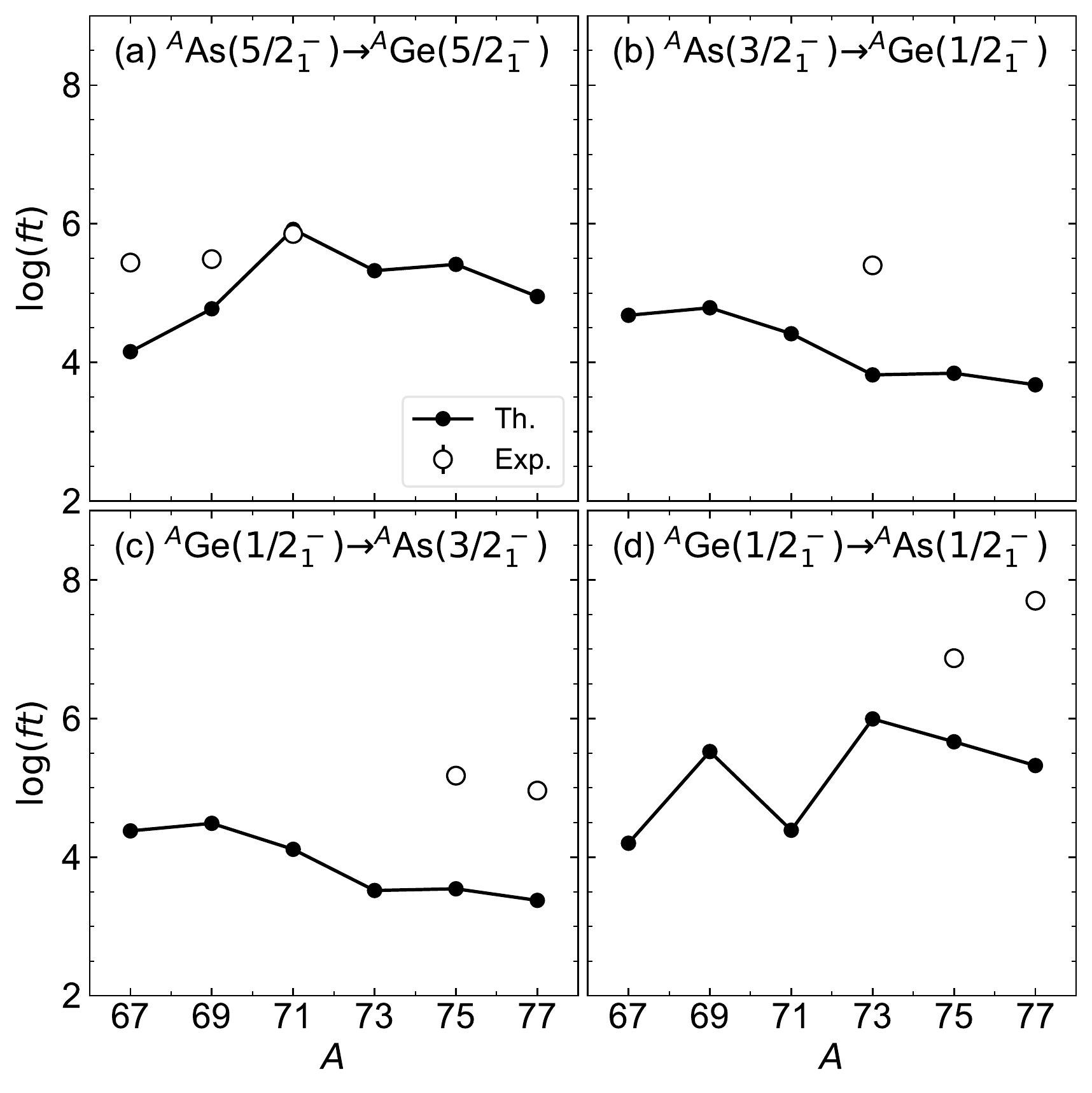}
\caption{Calculated $\ft$ values for the electron-capture 
decays from the odd-$A$ As to Ge nuclei, (a) $5/2^-_1\to5/2^-_1$ 
and (b) $3/2^-_1\to1/2^-_1$, and the $\btm$ decays from 
the odd-$A$ Ge to As nuclei, (c) $1/2^-_1\to3/2^-_1$ 
and (d) $1/2^-_1\to1/2^-_1$. The experimental data, 
available from Ref.~\cite{data}, are also shown.}
\label{fig:ft-odd}
\end{center}
\end{figure}

\subsection{$\beta$ decays between odd-$A$ nuclei\label{sec:bt-odd}}

Figure~\ref{fig:ft-odd} shows the calculated 
$\ft$ values for the electron-capture (EC) decays of the 
$5/2^-_1$ and $3/2^-$ states of the odd-$A$ 
As nuclei, and for the $\btm$ decays of the 
$1/2^-_1$ state of the odd-$A$ Ge nuclei. 
The behaviors of the predicted $\ft$ values reflect 
evolution of the underlying nuclear structure. 
A characteristic kink is observed at $A=71$ in the 
calculated $\ft$ values for 
the $\Delta I=0$ decays 
As$\,(5/2^-_1)\,\to\,$Ge$\,(5/2^-_1)\,$ and 
Ge$\,(1/2^-_1)\,\to\,$As$\,(1/2^-_1)\,$, 
shown in Figs.~\ref{fig:ft-odd}(a) 
and \ref{fig:ft-odd}(d), respectively. 
The $\ft$ results for the 
GT-type ($\Delta I=1$) decays 
As$\,(3/2^-_1)\,\to\,$Ge$\,(1/2^-_1)\,$ and 
Ge$\,(1/2^-_1)\,\to\,$As$\,(3/2^-_1)\,$ 
rather abruptly decrease from $A=71$ to 73. 

As shown in Fig.~\ref{fig:ft-odd}
the calculation systematically underestimates 
the observed $\ft$ values \cite{data}. 
For both the 
$^{67,69}$As$\,(5/2^-_1)\,\to\,^{67,69}$Ge$\,(5/2^-_1)\,$ 
decays, the dominant contributions to the GT and Fermi 
matrix elements come from the terms 
of the type $(a_{\nu f_{5/2}}^\+\times a_{\pi f_{5/2}})^{(I)}$, 
with $I=1$ and 0, respectively. 
As one can see in Figs.~\ref{fig:odd-wf}(e) and \ref{fig:odd-wf}(f), 
the ${5/2}^-_1$ states of both the parent ($^{67,69}$As) and 
daughter ($^{67,69}$Ge) nuclei are almost purely 
made of the $1f_{5/2}$ single-particle configurations, 
hence the GT and Fermi transitions 
are calculated to be so large as to give the 
small $\ft$ values compared to the data. 

For the GT-type decay 
$^{73}$As$\,(3/2^-_1)\,\to\,^{73}$Ge$\,(1/2^-_1)\,$, 
the largest contribution to the GT strength is from 
the term proportional to 
$s_\nu(a_{\nu p_{1/2}}^\+\times a_{\pi p_{3/2}})^{(1)}$. 
The IBFM-2 wave functions 
for the ${3/2}^-_1$ parent state of $^{73}$As and 
the ${1/2}^-_1$ daughter state of $^{73}$Ge are 
built predominantly on the $2p_{3/2}$ and 
$2p_{1/2}$ single-proton configurations, respectively 
[see Figs.~\ref{fig:odd-wf}(a) 
and \ref{fig:odd-wf}(d)]. Thus the GT strength 
between these states is calculated to be substantially 
large, hence the small $\ft$ value is obtained. 

The rapid increase of the $\ft$ values for 
As$\,(5/2^-_1)\,\to\,$Ge$\,(5/2^-_1)\,$ decay toward 
$A=71$ is attributed to the fact that the structure 
of the wave function for the ${5/2}^-_1$ daughter 
state drastically changes from $A=69$ to 71. 
One finds from 
Figs.~\ref{fig:odd-wf}(e) and \ref{fig:odd-wf}(f) 
that, the parent $^{71}$As$\,({5/2}^-_1)\,$ and 
the daughter $^{71}$Ge$\,({5/2}^-_1)\,$ states 
are here almost purely made of the proton $1f_{5/2}$  
and neutron $2p_{1/2}$ single-particle configurations, respectively.
Thus the terms such as  
$(a_{\nu f_{5/2}}^\+\times a_{\pi f_{5/2}})^{(I=0,1)}$, 
which make a large contribution to the 
GT and Fermi matrix elements for the decays of $^{67,69}$As, 
now play a much less important 
role in the case of the $^{71}$As decay. 
There are, instead, various other terms with 
small amplitudes in the GT and Fermi 
matrix elements of the $^{71}$As decay, 
which cancel each other and lead to 
the large $\ft$ value as compared to the $^{67,69}$As decays. 

As shown in 
Figs.~\ref{fig:ft-odd}(c) and \ref{fig:ft-odd}(d), 
the calculated $\ft$ values for the 
$\btm$ decays 
Ge$\,(1/2^-_1)\,\to\,$As$\,(3/2^-_1)\,$ and 
Ge$\,(1/2^-_1)\,\to\,$As$\,(1/2^-_1)\,$, 
show a certain systematic trend with $A$. 
Such behaviors also reflect 
the structure of the IBFM-2 
wave functions for the parent and daughter states. 

%
\begin{table}
\caption{\label{tab:ft-asge}
Comparison of calculated and observed \cite{data} 
$\ft$ values for the 
EC decays from odd-$A$ As to Ge nuclei.
}
 \begin{center}
 \begin{ruledtabular}
  \begin{tabular}{lccc}
& & \multicolumn{2}{c}{$\ft$} \\
\cline{3-4}
Decay & $I\to I'$ & Th. & Exp. \\
\hline
$^{67}$As$\to^{67}$Ge
& ${5/2}^{-}_{1}\to{5/2}^{-}_{1}$ & 4.15 & 5.44$\pm$0.13 \\
& ${5/2}^{-}_{1}\to{5/2}^{-}_{2}$ & 6.63 & 5.92$\pm$0.08\footnotemark[1] \\
& ${5/2}^{-}_{1}\to{5/2}^{-}_{3}$ & 6.08 & 6.4$\pm$0.4\footnotemark[1] \\
& ${5/2}^{-}_{1}\to{3/2}^{-}_{1}$ & 6.49 & 6.18$\pm$0.11\footnotemark[1] \\
& ${5/2}^{-}_{1}\to{3/2}^{-}_{2}$ & 7.61 & 5.64$\pm$0.07\footnotemark[1] \\
$^{69}$As$\to^{69}$Ge
& ${5/2}^{-}_{1}\to{5/2}^{-}_{1}$ & 4.77 & 5.49$\pm$0.02 \\
& ${5/2}^{-}_{1}\to{5/2}^{-}_{2}$ & 6.92 & 6.94$\pm$0.07 \\
& ${5/2}^{-}_{1}\to{5/2}^{-}_{3}$ & 5.63 & 6.80$\pm$0.06 \\
& ${5/2}^{-}_{1}\to{5/2}^{-}_{4}$ & 5.98 & 6.47$\pm$0.06 \\
& ${5/2}^{-}_{1}\to{5/2}^{-}_{5}$ & 7.15 & 5.95$\pm$0.05 \\
& ${5/2}^{-}_{1}\to{3/2}^{-}_{1}$ & 7.58 & 6.05$\pm$0.02 \\
& ${5/2}^{-}_{1}\to{3/2}^{-}_{2}$ & 7.44 & 7.21$\pm$0.05 \\
& ${5/2}^{-}_{1}\to{3/2}^{-}_{3}$ & 6.43 & 6.71$\pm$0.06 \\
& ${5/2}^{-}_{1}\to{3/2}^{-}_{4}$ & 7.07 & 5.82$\pm$0.05 \\
& ${5/2}^{-}_{1}\to{3/2}^{-}_{5}$ & 8.00 & 6.21$\pm$0.05 \\
& ${5/2}^{-}_{1}\to{7/2}^{-}_{1}$ & 10.85 & 6.20$\pm$0.05 \\
& ${5/2}^{-}_{1}\to{7/2}^{-}_{2}$ & 7.46 & 5.44$\pm$0.05 \\
$^{71}$As$\to^{71}$Ge
& ${5/2}^{-}_{1}\to{5/2}^{-}_{1}$ & 5.92 & 5.853$\pm$0.012 \\
& ${5/2}^{-}_{1}\to{5/2}^{-}_{2}$ & 6.28 & \\ 
& ${5/2}^{-}_{1}\to{5/2}^{-}_{3}$ & 6.55 & 6.869$\pm$0.015 \\
& ${5/2}^{-}_{1}\to{5/2}^{-}_{4}$ & 7.74 & 9.14$\pm$0.08\footnotemark[2] \\
& ${5/2}^{-}_{1}\to{5/2}^{-}_{5}$ & 7.30 & 6.840$\pm$0.025 \\
& ${5/2}^{-}_{1}\to{3/2}^{-}_{1}$ & 6.74 & 7.192$\pm$0.012 \\
& ${5/2}^{-}_{1}\to{3/2}^{-}_{2}$ & 7.47 & $>$8.6 \\
& ${5/2}^{-}_{1}\to{3/2}^{-}_{3}$ & 7.24 & 6.333$\pm$0.013 \\
& ${5/2}^{-}_{1}\to{3/2}^{-}_{4}$ & 8.25 & 7.430$\pm$0.023 \\
& ${5/2}^{-}_{1}\to{3/2}^{-}_{5}$ & 8.10 & 6.946$\pm$0.014 \\
& ${5/2}^{-}_{1}\to{7/2}^{-}_{1}$ & 8.38 & 8.79$\pm$0.25 \\
& ${5/2}^{-}_{1}\to{7/2}^{-}_{2}$ & 7.85 & 7.296$\pm$0.016 \\
$^{73}$As$\to^{73}$Ge
& ${3/2}^{-}_{1}\to{1/2}^{-}_{1}$ & 3.82 & 5.4 \\
 \end{tabular}
\footnotetext[1]{Parity not firmly established.}
\footnotetext[2]{$I=({3/2},{5/2}^-)$ level at 886 keV}
 \end{ruledtabular}
 \end{center}
\end{table}

%
\begin{table}
\caption{\label{tab:ft-geas}
Same as Table~\ref{tab:ft-asge}, but for the $\btm$ 
decays from odd-$A$ Ge to As nuclei.
}
 \begin{center}
 \begin{ruledtabular}
  \begin{tabular}{lccc}
& & \multicolumn{2}{c}{$\ft$} \\
\cline{3-4}
Decay & $I\to I'$ & Th. & Exp. \\
\hline
$^{75}$Ge$\to^{75}$As
& ${1/2}^{-}_{1}\to{3/2}^{-}_{1}$ & 3.54 & 5.175$\pm$0.007 \\
& ${1/2}^{-}_{1}\to{3/2}^{-}_{2}$ & 5.04 & 5.63$\pm$0.05 \\
& ${1/2}^{-}_{1}\to{3/2}^{-}_{3}$ & 5.90 & 6.42$\pm$0.06\footnotemark[1] \\
& ${1/2}^{-}_{1}\to{1/2}^{-}_{1}$ & 5.66 & 6.87$\pm$0.05 \\
& ${1/2}^{-}_{1}\to{1/2}^{-}_{2}$ & 4.45 & 6.94$\pm$0.05 \\
& ${1/2}^{-}_{1}\to{1/2}^{-}_{3}$ & 7.18 & 6.42$\pm$0.06\footnotemark[1] \\
$^{77}$Ge$\to^{77}$As
& ${1/2}^{-}_{1}\to{3/2}^{-}_{1}$ & 3.38 & 4.96$\pm$0.04 \\
& ${1/2}^{-}_{1}\to{3/2}^{-}_{2}$ & 3.79 & 7.2$\pm$0.2 \\
& ${1/2}^{-}_{1}\to{3/2}^{-}_{3}$ & 4.71 & 5.3$\pm$0.1 \\
& ${1/2}^{-}_{1}\to{3/2}^{-}_{4}$ & 4.82 & 7.2$\pm$0.1 \\
& ${1/2}^{-}_{1}\to{3/2}^{-}_{5}$ & 5.15 & 5.7$\pm$0.1\footnotemark[2] \\
& ${1/2}^{-}_{1}\to{1/2}^{-}_{1}$ & 5.32 & 7.7$\pm$0.1 \\
& ${1/2}^{-}_{1}\to{1/2}^{-}_{2}$ & 4.91 & 5.7$\pm$0.1\footnotemark[2] \\
& ${1/2}^{-}_{1}\to{1/2}^{-}_{3}$ & 3.98 & 5.8$\pm$0.1\footnotemark[3] \\
  \end{tabular}
\footnotetext[1]{$I={1/2}^-$ or ${3/2}^-$ state at 618 keV}
\footnotetext[2]{$I={1/2}^-$ or ${3/2}^-$ state at 1605 keV}
\footnotetext[3]{$I={1/2}^-$ or ${3/2}^-$ state at 1676 keV}
 \end{ruledtabular}
 \end{center}
\end{table}

For the sake of completeness, Tables~\ref{tab:ft-asge} and 
\ref{tab:ft-geas} show the calculated 
and experimental $\ft$ values for those $\beta$ decays 
for which the data are available. 

One can also make a comparison with the previous 
IBFM-2 calculation for the $^{69,71}$As $\to$ $^{69,71}$Ge 
$\beta$ decays in Ref.~\cite{brant2004}. 
In general, the $\ft$ values obtained in the present 
study appear to be systematically larger than those 
reported in Ref.~\cite{brant2004}. 
Particularly for the decay 
$^{69}$As$\,(5/2^-_1)\,\to\,^{69}$Ge$\,(3/2^-_1)\,$, 
here the value $\ft=7.58$ is obtained, 
which overestimates the experimental one, $6.05\pm0.02$.
On the other hand, 
the calculation in Ref.~\cite{brant2004} gave a smaller 
value $\ft=5.88$, in a better agreement with experiment. 
The $\ft$ values for the $\Delta I=0$ decay 
$^{69}$As$\,(5/2^-_1)\,\to\,^{69}$Ge$\,(5/2^-_1)\,$ 
obtained from both calculations are close to each 
other, i.e., $\ft=4.26$ in Ref.~\cite{brant2004} and the present 
one is slightly larger, $\ft=4.77$. 
Both calculations underestimate the data, $\ft=5.49\pm0.02$. 
For both IBFM-2 calculations, the results for the $\ft$ values 
appear to be quite sensitive 
to the wave functions for the initial and final odd-$A$ 
nuclei. In fact, the IBFM-2 framework in Ref.~\cite{brant2004} 
is largely based on the phenomenological grounds, i.e., 
the empirical parameters for the even-even IBM-2 core Hamiltonian, 
and the phenomenological single-particle energies 
were adopted in that reference, 
while most of the parameters are here determined based on 
the EDF calculations.

To make a reasonable comparison with the experimental 
$\ft$ data, one could estimate the effective value of the 
$\ga/\gv$ ratio for the GT matrix element 
[see Eq.~(\ref{eq:ft})], denoted here by $\ave$. 
By fitting to the experimental $\ft$ value 
for the $\Delta I=1$ $\beta$-decay 
$^{73}$As$\,(3/2^-_1)\,\to\,^{73}$Ge$\,(1/2^-_1)\,$ 
one obtains $\ave=0.206$, 
equivalent to a quenching factor $q=0.162$. 
If one applies this $\ave$ value, for instance, to the 
$\Delta I=0$ decay 
$^{69}$As$\,(5/2^-_1)\,\to\,^{69}$Ge$\,(5/2^-_1)\,$, 
then the $\ft$ value is only slightly 
increased to $\ft=5.06$, with respect to the one (4.77)
obtained with the free $\ga/\gv$ ratio.  
Likewise, the $\ave$ values for the 
$^{75}$Ge$\,(1/2^-_1)\,\to\,^{75}$As$\,(3/2^-_1)\,$ and 
$^{75}$Ge$\,(1/2^-_1)\,\to\,^{75}$As$\,(3/2^-_1)\,$ decays 
are calculated to be 0.194 and 0.204, respectively. 
Also in these cases, the quenching of the $\ga/\gv$ ratio 
does not drastically increase the relevant $\Delta I=0$ 
$\beta$ decay, i.e., the $1/2^-_1\,\to\,1/2^-_1$ one. 
%
%
%

\begin{figure}[ht]
\begin{center}
\includegraphics[width=\linewidth]{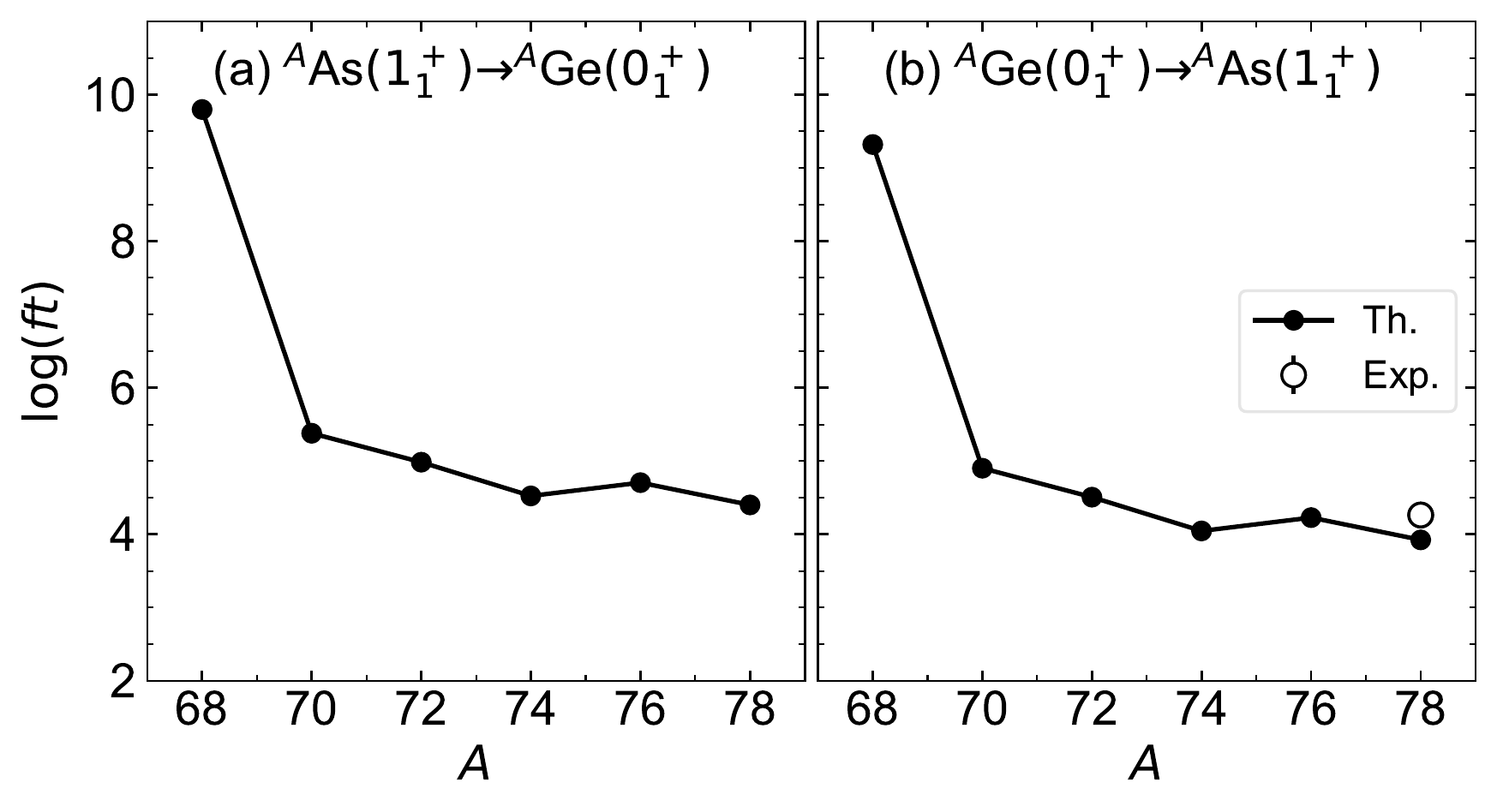}
\caption{Same as Fig.~\ref{fig:ft-odd}, but for the 
$\btp$/EC decay $1^+_1\to0^+_1$ from the even-$A$ As to Ge nuclei (a), 
and the $\btm$ decay $0^+_1\to1^+_1$ from the even-$A$ Ge to As 
nuclei (b).}
\label{fig:ft-even}
\end{center}
\end{figure}

\subsection{$\beta$ decays between even-$A$ nuclei\label{sec:bt-even}}

In Fig.~\ref{fig:ft-even} shown are  
the calculated $\ft$ values for the GT-type transitions 
between the even-$A$ nuclei, i.e., 
As$\,(1^+_1)\,\to\,$Ge$\,(0^+_1)\,$ $\btp$/EC and 
Ge$\,(0^+_1)\,\to\,$As$\,(1^+_1)\,$ $\btm$ decays. 
The present calculation gives the values typically within 
the range $4\lesssim\ft\lesssim5$ for both the $\btp$/EC 
and $\btm$ decays. 
Apart from the decays of $^{68}$As and $^{68}$Ge, 
the predicted $\ft$ values show a gradual decrease with 
$A$, but with a kink at $A=76$. 
Note that a good agreement with the observed $\ft$ value 
is seen for the $\btm$ decay 
$^{78}$Ge$\,(0^+_1)\,\to\,^{78}$As$\,(1^+_1)\,$ 
(see Fig.~\ref{fig:ft-even}(b)). 
As shown in Sec.~\ref{sec:doo}, the IBFFM-2 wave 
functions for the $1^+_1$ state of those odd-odd As 
nuclei with $A\geqslant72$ are dominated by the 
neutron-proton pair configuration 
$[\nu{p_{1/2}}\otimes\pi{p_{3/2}}]^{J=1^+}$. 
For the GT transitions between those nuclei with $A\geqslant72$, 
components in the GT operator of the forms 
proportional to 
$(a_{\nu p_{1/2}}^{\+}\times a_{\pi p_{3/2}}^\+)^{(1)}$, 
$(a_{\nu p_{3/2}}^{\+}\times a_{\pi p_{1/2}}^\+)^{(1)}$, and 
$(a_{\nu p_{3/2}}^{\+}\times a_{\pi p_{3/2}}^\+)^{(1)}$
have particularly large contributions to the 
$\beta$-decay rates.


On the other hand, 
as mentioned above, the calculation predicts 
exceptionally large $\ft$ values ($\approx9-10$) 
for the decays of the states 
$^{68}$As$\,(1^+_1)\,$ and $^{68}$Ge$\,(0^+_1)\,$. 
This is, to a large extent, traced back to the 
IBFFM-2 wave function of the $1^+_1$ state of the 
$^{68}$As nucleus, in which 
the fraction of the component 
$[\nu{p_{1/2}}\otimes\pi{p_{3/2}}]^{J=1^+}$ 
is negligibly small, in comparison to 
the decays of the $A\geqslant72$ nuclei. 
In the GT matrix elements, 
the terms corresponding to the coupling between 
the neutron $2p_{1/2}$ (or $2p_{3/2}$) 
and proton $2p_{3/2}$ (or $2p_{1/2}$)
single-particle states 
make only vanishing contributions, 
while other terms have 
small amplitudes and cancel each other.

Table~\ref{tab:ft-even} shows a comparison between the 
calculated and experimental data for the $\beta$-decay 
$\ft$ values for the ground states 
of even-$A$ nuclei. 
The predicted $\ft$ values 
for the relevant $\beta$ decays, i.e., 
$^{68}$As$\,(3^+_1)\,\to\,^{68}$Ge$\,(2^+_1)\,$, 
$^{68}$As$\,(3^+_1)\,\to\,^{68}$Ge$\,(4^+_1)\,$, 
$^{70}$As$\,(4^+_1)\,\to\,^{70}$Ge$\,(4^+_1)\,$, and 
$^{78}$Ge$\,(0^+_1)\,\to\,^{78}$Ge$\,(1^+_1)\,$, 
are generally in a fair agreement with the observed 
ones. 
The IBFFM-2 $\ft$ values for the 
$^{70}$As$\,(4^+_1)\,\to\,^{70}$Ge$\,(3^+_1)\,$ decay 
is nearly twice as large as the observed one. 
The large deviation is probably related to the fact that 
the IBFFM-2 calculation does not give the correct 
ground-state spin of $I=4^+$ for $^{70}$As. 
For this particular decay, 
numerous terms are fragmented and cancel each other 
in the GT matrix element, 
giving rise to the too large $\ft$ value. 

%
\begin{table}
\caption{\label{tab:ft-even}
Same as Table~\ref{tab:ft-asge}, but for the $\btp$/EC decays 
from even-$A$ As to Ge nuclei, and for the $\btm$ decays 
from $^{78}$Ge to $^{78}$As. 
}
 \begin{center}
 \begin{ruledtabular}
  \begin{tabular}{lccc}
& & \multicolumn{2}{c}{$\ft$} \\
\cline{3-4}
Decay & $I\to I'$ & Th. & Exp. \\
\hline
$^{68}$As$\to^{68}$Ge
& $3^{+}_{1}\to2^{+}_{1}$ & 6.66 & 7.38$\pm$0.24 \\
& $3^{+}_{1}\to2^{+}_{2}$ & 6.95 & 6.86$\pm$0.19 \\
& $3^{+}_{1}\to2^{+}_{3}$ & 6.34 & 6.89$\pm$0.10 \\
& $3^{+}_{1}\to2^{+}_{4}$ & 5.81 & 7.24$\pm$0.04 \\
& $3^{+}_{1}\to2^{+}_{5}$ & 7.21 & 6.57$\pm$0.04 \\
& $3^{+}_{1}\to4^{+}_{1}$ & 6.34 & 7.02$\pm$0.06 \\
& $3^{+}_{1}\to4^{+}_{2}$ & 5.73 & 6.74$\pm$0.03 \\
& $3^{+}_{1}\to4^{+}_{3}$ & 6.63 & 5.979$\pm$0.018\footnotemark[1] \\
$^{70}$As$\to^{70}$Ge
& $4^{+}_{1}\to4^{+}_{1}$ & 6.58 & 7.30$\pm$0.16 \\
& $4^{+}_{1}\to4^{+}_{2}$ & 6.03 & 7.37$\pm$0.14 \\
& $4^{+}_{1}\to4^{+}_{3}$ & 6.01 & 5.69$\pm$0.05 \\
& $4^{+}_{1}\to3^{+}_{1}$ & 10.74 & 6.97$\pm$0.04 \\
$^{78}$Ge$\to^{78}$As
& $0^{+}_{1}\to1^{+}_{1}$ & 3.92 & 4.264$\pm$0.025 \\
& $0^{+}_{1}\to1^{+}_{2}$ & 5.15 & 5.61$\pm$0.12 \\
  \end{tabular}
\footnotetext[1]{Level at 3042 keV with spin and parity temporarily assigned to be $({4^+})$.}
 \end{ruledtabular}
 \end{center}
\end{table}

By following the same procedure as discussed in the 
previous section, the $\ave$ ratios can be extracted 
for the $\beta$ decays of the even-$A$ nuclei. 
Of particular interest is the 
$^{78}$Ge$\,(0^+_1)\,\to\,^{78}$As$\,(1^+_1)\,$ $\btm$ decay, 
since the neighboring nucleus $^{76}$Ge is a 
candidate for the $\znbb$-decay emitter. 
For the above decay, one obtains the effective 
ratio $\ave=0.860$, corresponding to the 
quenching factor $q=0.677$. 
This appears to be a modest value, 
as compared with those obtained here for the 
$\beta$ decays of odd-$A$ nuclei. 
The result is also more or less consistent with a common 
value of the effective $\ave$ ratio that is often considered 
in many of the calculations for the single-$\beta$ and $\db$ 
decays of $^{76}$Ge \cite{suhonen2017}. 
One also obtains the $\ave$ values for the $^{68}$As decays. 
For instance, the value $\ave=0.554$ is extracted 
for the $3^+_1\to2^+_1$ decay. 

\section{Concluding remarks\label{sec:summary}}

The shape evolution and the related 
spectroscopic properties of the low-lying states, and 
the $\beta$-decay properties of the 
even-even, odd-mass, and odd-odd Ge and As nuclei 
in the mass $A\approx70-80$ region have been investigated 
within the framework of the nuclear EDF and the particle-boson 
coupling scheme. 
The constrained SCMF calculation based on the universal 
relativistic functional DD-PC1 and the separable pairing 
force of finite range provides triaxial quadrupole 
deformation energy surface for the even-even $^{66-78}$Ge 
nuclei. By mapping the mean-field energy surface onto 
the expectation value of the IBM-2 Hamiltonian in 
the coherent state, the parameters for the Hamiltonian 
have been determined. The same SCMF calculation yields 
spherical single(quasi)-particle energies and occupation 
probabilities, which are the essential building blocks 
of the particle-boson interactions, and the Gamow-Teller 
and Fermi  transition operators. 
Fixed values are employed for the three coupling 
constants for the particle-boson 
interaction terms, and for the two parameters for the 
residual neutron-proton interaction in the IBFFM-2, 
which are determined to have an overall reasonable 
agreement with the low-energy data for the odd-$A$ and 
odd-odd nuclei under study.

At the SCMF level, a rapid nuclear structural evolution 
as a function of the nucleon number has been suggested 
in the energy surface, from the $\gamma$-soft oblate shapes 
for $^{66-70}$Ge, to the spherical-oblate shape coexistence 
for $^{72}$Ge, to the triaxial deformation for $^{74}$Ge, 
and to the $\gamma$-soft prolate shapes for $^{76,78}$Ge. 
The resultant energy spectra for the low-lying states, 
obtained by the diagonalization of the mapped IBM-2 
Hamiltonian, follow the observed systematics with $A$. 
The possibility of shape coexistence 
for $^{72}$Ge has been addressed, in which nucleus the observed 
spectrum is characterized by the 
low-lying excited $0^+$ state below the $2^+_1$ one. 
The configuration-mixing IBM-2 calculation reproduces 
well the observed $0^+_2$ excitation energy. 
The calculated low-lying 
negative-parity levels in the neighboring odd-$N$ Ge 
and odd-$Z$ As, as well as the odd-odd As, 
nuclei show the systematic behaviors reflecting 
the shape transition that is suggested to occur in the 
even-even Ge core. The $B(E2)$ and $B(M1)$ transition 
rates in the odd-nucleon systems are, however, sensitive 
to the IBFM-2 or IBFFM-2 wave functions. 

The wave functions for the even- and odd-$A$ nuclei 
have been then used to compute the GT and Fermi matrix 
elements for the $\beta$ decays. 
The predicted $\ft$ values for the 
$\beta$ decays of the odd-$A$ nuclei evolve with $A$, 
corroborating the underlying nuclear structure evolution 
in the parent and daughter nuclear systems. 
As compared to experiment, the $\ft$ values obtained 
for the odd-$A$ nuclei are systematically 
small. This reflects the nature of the corresponding 
IBFM-2 wave functions and, in turn, serves as a 
sensitive test of 
various model assumptions and the microscopic input 
provided by the EDF. To effectively account for the 
deviation between the IBFM-2 and observed $\ft$ values, 
drastic quenching for the ratio $\ga/\gv$ would be required 
for the odd-$A$ cases. 
For the decays of the even-$A$ As and Ge nuclei, 
the agreement with the experimental $\ft$ values has turned 
out to be slightly better, and rather modest quenching has been 
suggested. 

The present theoretical scheme allows for a simultaneous 
and computationally feasible calculation of the low-lying 
states and their $\beta$ decays for all kinds of nuclei, 
i.e., even-even, odd-mass, and odd-odd ones,  
based largely on the nuclear EDF calculations. 
The next step is the applications of the methodology to 
the $\beta$-decay properties of more neutron-rich nuclei, 
which are expected to play a significant role in the astrophysical 
processes and are experimentally of much interest. 
On the other hand, the reported spectroscopic calculation 
in this mass region opens up 
a possibility to study the effects of the shape coexistence, 
as well as the triaxial deformation,  
on the low-lying states of the odd-nucleon systems within 
the IBFM-2 and IBFFM-2, and on the predictions on their 
$\beta$ decays. These would require a major extension 
of the model. 
Work along these lines is in progress, and will be 
reported elsewhere.

\acknowledgments
This work is financed within the Tenure Track Pilot Programme of 
the Croatian Science Foundation and the \'Ecole Polytechnique 
F\'ed\'erale de Lausanne, and Project No. TTP-2018-07-3554 Exotic 
Nuclear Structure and Dynamics, with funds of the Croatian-Swiss 
Research Programme.

\bibliography{refs}

\end{document}